\documentstyle[11pt,epsf,doublespace]{article}
\newcommand{\et}{{\rm E}_{\scriptscriptstyle\rm T}}

\newcommand{\met}{\mbox{$\protect \raisebox{.3ex}{$\not$}\et$}}

\newcommand{\bbbar} { {\rm b} \bar{{\rm b}}  }

\newcommand{\be}{\begin{equation}}
\newcommand{\ee}{\end{equation}}
\newcommand{\TTbar}{\mbox{$t\overline{t}$}}

 1

\begin{document}
\begin{flushright}
FERMILAB-PUB-94/411-E\\
December 12, 1994\\
\end{flushright}
\begin{center}
\begin{large}
{\bf Kinematic Evidence for Top Quark Pair Production in \\
 $W+$ Multijet Events in $p \bar p$ Collisions at $\sqrt{s}$~=~1.8 TeV}
\end{large}
\vspace{2ex}
\font\eightit=cmti8
\def\r#1{\ignorespaces $^{#1}$}
\hfilneg
\begin{sloppypar}
\noindent
F.~Abe,\r {13} M.~G.~Albrow,\r 7 S.~R.~Amendolia,\r {23} D.~Amidei,\r {16}
J.~Antos,\r {28} C.~Anway-Wiese,\r 4 G.~Apollinari,\r {26} H.~Areti,\r 7
M.~Atac,\r 7 P.~Auchincloss,\r {25}
% ...
\end{sloppypar}

\vskip .025in
\begin{center}
(CDF Collaboration)
\end{center}

\vskip .025in
\begin{center}
\r 1  {\eightit Argonne National Laboratory, Argonne, Illinois 60439} \\
\r 2  {\eightit Istituto Nazionale di Fisica Nucleare, University of Bologna,
I-40126 Bologna, Italy} \\
\r 3  {\eightit Brandeis University, Waltham, Massachusetts 02254} \\
\r 4  {\eightit University of California at Los Angeles, Los
Angeles, California  90024} \\
\r 5  {\eightit University of Chicago, Chicago, Illinois 60637} \\
\r 6  {\eightit Duke University, Durham, North Carolina  27708} \\
\r 7  {\eightit Fermi National Accelerator Laboratory, Batavia, Illinois
60510} \\
\r 8  {\eightit Laboratori Nazionali di Frascati, Istituto Nazionale di Fisica
               Nucleare, I-00044 Frascati, Italy} \\
\r 9  {\eightit Harvard University, Cambridge, Massachusetts 02138} \\
\r {10} {\eightit University of Illinois, Urbana, Illinois 61801} \\
\r {11} {\eightit Institute of Particle Physics, McGill University, Montreal
H3A 2T8, and University of Toronto,\\ Toronto M5S 1A7, Canada} \\
\r {12} {\eightit The Johns Hopkins University, Baltimore, Maryland 21218} \\
\r {13} {\eightit National Laboratory for High Energy Physics (KEK), Tsukuba,
Ibaraki 305, Japan} \\
\r {14} {\eightit Lawrence Berkeley Laboratory, Berkeley, California 94720} \\
\r {15} {\eightit Massachusetts Institute of Technology, Cambridge,
Massachusetts  02139} \\
\r {16} {\eightit University of Michigan, Ann Arbor, Michigan 48109} \\
\r {17} {\eightit Michigan State University, East Lansing, Michigan  48824} \\
\r {18} {\eightit University of New Mexico, Albuquerque, New Mexico 87131} \\
\r {19} {\eightit Osaka City University, Osaka 588, Japan} \\
\r {20} {\eightit Universita di Padova, Instituto Nazionale di Fisica
          Nucleare, Sezione di Padova, I-35131 Padova, Italy} \\
\r {21} {\eightit University of Pennsylvania, Philadelphia,
        Pennsylvania 19104} \\
\r {22} {\eightit University of Pittsburgh, Pittsburgh, Pennsylvania 15260} \\
\r {23} {\eightit Istituto Nazionale di Fisica Nucleare, University and Scuola
               Normale Superiore of Pisa, I-56100 Pisa, Italy} \\
\r {24} {\eightit Purdue University, West Lafayette, Indiana 47907} \\
\r {25} {\eightit University of Rochester, Rochester, New York 14627} \\
\r {26} {\eightit Rockefeller University, New York, New York 10021} \\
\r {27} {\eightit Rutgers University, Piscataway, New Jersey 08854} \\
\r {28} {\eightit Academia Sinica, Taiwan 11529, Republic of China} \\
\r {29} {\eightit Superconducting Super Collider Laboratory, Dallas,
Texas 75237} \\
\r {30} {\eightit Texas A\&M University, College Station, Texas 77843} \\
\r {31} {\eightit Texas Tech University, Lubbock, Texas 79409} \\
\r {32} {\eightit University of Tsukuba, Tsukuba, Ibaraki 305, Japan} \\
\r {33} {\eightit Tufts University, Medford, Massachusetts 02155} \\
\r {34} {\eightit University of Wisconsin, Madison, Wisconsin 53706} \\
\r {35} {\eightit Yale University, New Haven, Connecticut 06511} \\
\end{center}
\newpage
\begin{abstract}
We present a study of $W+$multijet events that compares the
 kinematics of the observed events with expectations from direct QCD $W+$jet
production and from production and decay of top quark pairs.
The data were collected in the 1992-93 run
with the Collider Detector at Fermilab (CDF)
 from 19.3 pb$^{-1}$ of proton-antiproton collisions at $\sqrt{s} = 1.8$ TeV.
A $W + \geq 2$ jet sample and a $W + \geq 3$ jet sample are
selected with the requirement that at least the two or three jets
 have energy transverse with respect to the beam axis in excess of 20~GeV.
 The jet energy distributions for the $W + \geq 2$ jet sample agree well
with the predictions of direct QCD $W$ production.
{}From the $W + \geq 3$ jet events,
a ``signal sample" with an improved ratio of $t \bar{t}$ to QCD produced $W$
events is
selected by requiring each jet to be emitted centrally in the event center
of mass frame.  This sample contains 14 events with unusually hard
 jet E$_T$ distributions not well described by expectations for jets
from direct QCD $W$ production and other background processes.  Using expected
jet E$_T$ distributions, a relative likelihood is defined and used to determine
if an event is more consistent with the decay of $t \bar{t}$ pairs, with
M$_{top}$~=~170 GeV/c$^2$, than with direct QCD $W$ production.
Eight of the 14 signal sample
events are found to be more consistent with top than direct QCD $W$ production,
while only 1.7 such top--like events are expected in the absence of
$t \bar{t}$.  The
probability that the observation is due to an upward fluctuation of the number
of background events is found to be 0.8\%.  The robustness of the
result was tested by varying the cuts defining the signal sample, and the
largest probability for such a fluctuation found was 1.9\% .
Good agreement in the jet spectra is obtained if
 jet production from  $t \bar{t}$ pair decays is included.
For those events kinematically more consistent with $t \bar{t}$
we
find evidence for a $b$--quark content in their jets
to the extent expected from top decay, and larger than expected for
background processes. For events with four or more jets,  the discrepancy
with the predicted jet energy distributions from direct QCD $W$ production,
and the associated excess of $b$--quark content is more pronounced.
\end{abstract}
\end{center}
\newpage
\section{Introduction}

\hspace{0.5cm} Recently CDF presented evidence for top quark pair
production, both via the observation of events with two high $P_T$ leptons and
via the observation of events with a $W$, 3 or more jets, and a jet tagged as a
$b$ quark~\cite{prd}.
In that analysis the distributions of specific
{\it kinematic} parameters of the events, such as jet energies and
angles,
were not used to discriminate between signal and background.
It is of interest to search for
 evidence of top quark pair production based on this event structure and
 to determine whether one can select a top quark enriched sample of events
with suitable cuts on kinematic variables. \\
\indent The main background to $t\bar t$ production comes
from higher order QCD production of
quarks and gluons in association with direct $W$ production.
 Recent experiments have indicated that the top quark
mass is larger than of order
130 GeV/c$^{2}$~\cite{prd}~\cite{D0}~\cite{lep}.
For such a heavy top quark it is difficult
to distinguish the signal from the background based on the properties of
the $W$.
 However,
the jets in $t\bar{t}$ events
have higher energies on average than those accompanying the $W$ in
direct $W$ production, and
are expected to be produced at larger angles relative to the beam
direction.
In this paper we
separate $t \bar t$ events where one of the $W$'s decays leptonically and the
other hadronically
($t \bar{t} \rightarrow W^+b + W^- \bar{b} \rightarrow l \nu b +
q\bar{q} \bar{b}$) by exploiting these properties. \\
\indent The paper is organized as follows. Section 2 describes the analysis
 cuts used to define the $W +$ jet sample. Section 3 gives a brief
comparison of the kinematics of directly produced
 $W +$ jet events and of $W +$ jet events
from top quark decay, and explains the analysis strategy.
 Section 4 summarizes various comparisons of QCD Monte Carlo
predictions which successfully fit experimental measurements in
processes where a top quark contribution can be neglected.
 Kinematic features of our $W + \geq $ 3
 jet data sample are compared to background  and to
top quark prediction in Section 5.
This comparison shows evidence for a top quark--like
component in the data.
 Section 6 combines this
 result with the independent information obtained from the
algorithms which provide identification of  $b$ quarks in the events.
 In Section 7, as an additional test, we look  for an excess of $W +4$ jet
events in the top quark candidate sample. The conclusions
are presented in Section 8.
\section{Data Selection}

\hspace{0.5cm} This analysis is based on 19.3 pb$^{-1}$ of data from 1.8 TeV
$p\overline{p}$ collisions taken with the CDF detector during the
1992-1993 Tevatron run.
The CDF detector is described in detail elsewhere
{}~\cite{prd},\cite{4},\cite{5}.
For this run, the tracking system was upgraded with a high precision silicon
vertex detector (SVX)~\cite{5}), and the muon detector was improved
at pseudorapidity~\cite{eta}
  $|\eta|$$<$0.6 by adding an absorber of 0.6 m of
steel followed by drift chambers.
In addition, the coverage of the central  muon detector was
extended to the region of pseudorapidity
$0.6<|\eta |<1.0$ (over about 2/3 of the azimuth)
with drift chambers and scintillation counters.
 The transverse momentum P$_T$,  defined as P$_T$~=~P$\sin\theta$,
is the projection of the observed momentum (P) onto the
plane transverse to the beam axis.
Similarly, the transverse energy is defined as:
E$_T$= E$\sin\theta$, where E is the energy measured in the calorimeter.
The identification and measurement of isolated, high--P$_T$
electrons and  muons,
the measurement of the missing E$_T$ ($\met$)  indicative of
neutrinos in the events,  the jet clustering algorithm, and the jet energy
corrections are discussed in Ref.~\cite{prd},\cite{3jetprd} and
\cite{4jetprd}.\\
\indent   A sample of $W \rightarrow e\nu (\mu\nu$) candidate
events was selected with the requirement
 that E$^e_T$$>$20 GeV (P$^{\mu}_T$$>$20 GeV/c) and  $\met$$>$25 GeV.
 In addition,
the transverse mass,
defined as M$_T$ = [2E$_T$$\met$(1-cos$\Delta\phi$)]$^{1/2}$ (where
$\Delta\phi$ is the difference in azimuthal angle between the
missing energy direction and the lepton), was required to be
larger than 40 GeV/c$^2$.
   The jets are
reconstructed with a cone size $R$ = $\sqrt{\Delta\Phi^{2} +
\Delta\eta^{2}}$ = 0.4
(where $\Delta\Phi$ is the cone half--width in azimuth and
$\Delta\eta$ is the cone half--width in pseudorapidity).
The jet  energies are
corrected by a rapidity and energy dependent factor which  accounts
for calorimeter non-linearity and reduced response at detector
boundaries~\cite{3jetprd}, \cite{4jetprd}.
In addition to these detector effects, a correction is also made
for energy which is radiated out of the jet reconstruction cone.
The  $\met$ is calculated after correcting the jet
energies.
In order to allow for a clean separation of jets from each other and to
facilitate the comparison of energy distributions with theoretical
expectations, the centroids of the three leading jets are required to be
separated from each other  by  $\Delta R \geq 0.7$. \\
\indent Backgrounds from $Z$ decay, Drell Yan production of dileptons, and
possible $t\bar t$ events in which both $W$'s decay leptonically  are
removed by  rejecting events with an additional isolated
track with P$_T$ $>$ 15 GeV/c in the
central tracking system that is not associated with the primary
lepton.
Tracks are defined as isolated when the total
transverse momentum of the charged tracks (other than the track in question)
in a cone of radius $R$ = 0.4 centered on the track is less than 0.1 times the
P$_T$ of the  track.
A study of a QCD multijet sample has shown
that fewer than 1\%
of jets with E$_T$(jet) $>$ 20 GeV are rejected by this cut \cite{sandra}.
An additional $Z$ removal
algorithm eliminates events with an oppositely--charged dilepton ($ee$
or $\mu\mu$) invariant mass in the range 70 to 110 GeV/c$^2$.\\
\indent A sample of $W + \geq$  2 jets with the two
leading jets having E$_T$(jet) $>$ 20 GeV and
$|$$\eta$(jet)$|$ $<$ 2  is selected, where the $t \bar t$ contribution is
expected to be relatively small. This sample is studied in order to check
 whether the energy spectra of the leading jets agree with QCD prediction
 for direct   $W +$ jet production. The search for a $t \bar{t}$ component
 is performed in the subsample with $\geq$ 3 such jets.\\
\indent The primary differences in event selection between this analysis
and the analysis of $W + \geq 3$ jets performed in Ref.~\cite{prd}
 are that (1) corrections to jet energy and $\met$ are
made {\it prior} to event selection, (2) the jets are explicitly required
to be separated by $\Delta R > 0.7$, (3) a cut is added on the transverse mass,
and (4) the rejection of events with an additional isolated track
 is included. These changes are made
to simplify the comparison of observed jet energies with
theoretical predictions for direct $W +$ jet production and
to reduce background from non-$W$ sources. \\
\indent The fraction of all $t \bar t$ events that should fall into
our $W + \geq 3$ jet
sample is determined from the HERWIG
$p \bar p \rightarrow t \bar t$ Monte Carlo program and the CDF detector
simulation. Corrections to the acceptance for trigger inefficiencies  and
differences in lepton identification between data and Monte Carlo simulation
 are identical to those described in Ref.~\cite{prd}. For this analysis
we find that the top quark acceptance ranges from 2.7$\pm$0.2\% at
M$_{top}$=150 GeV/c$^2$ to 3.0$\pm$0.2\% at M$_{top}$=190 GeV/c$^2$.
 The number of $t \bar t$ events expected in the $W + \geq 3$ jet sample
using the Standard Model top quark production cross section from
Ref.~\cite{laenen} is about 7 for M$_{top}$ = 170 GeV/c$^2$. Using the
 cross section from Ref.~\cite{prd} we expect  $16.7^{+7.6}_{-6.0}$ events.
We observe a total of 49 events, 25 of them being in common with the
52 $W +$ $\ge$ 3 jet event sample of Ref.~\cite{prd}.  For top quark events
the two sets of cuts make little difference: about 90\% of all top quark events
contained in the sample of 49 events will also show up in the sample of 52.
 QCD $W +$ jet events often will be close to the jet E$_T$ cuts. Therefore
only approximately 67\% of QCD events from the sample of 49 will also
be found under the cuts of the sample of 52. If we assume that all 49
events are from QCD, then we expect  an overlap of approximately 33
events to be compared to the observed overlap of 25 events.
\section{$W$ and $t \bar{t}$ Kinematics}

\hspace{0.5cm} Monte Carlo event samples are used to compare the
distribution of several kinematic variables for top quark and background
events.
Samples of top quark events of
various masses were generated with both ISAJET~\cite{9} and
 HERWIG~\cite{HERWIG}, and it was verified that both Monte Carlo generators
give similar results. $W+$ jet events
 were  generated according to the lowest order matrix elements for the
production of a  $W$ with $n$ final state partons. The complete sets of matrix
elements at tree level have been determined for $n=0,1,2,3,4$ and  are
implemented in the program VECBOS~\cite{7}.
 To avoid infrared divergences
which would occur at small angles and small $P_T$, cuts are applied in the
event generation that require $P_T$ (parton) $>$ 10 GeV/c,
$|$$\eta$ (parton)$|$ $<$ 3.5, and $\Delta R$ (parton-parton) $>$ 0.4.
 Unless otherwise noted,
 $Q^2$= M$_W^2$ has been used  for the  $\alpha_s$ scale and
the structure functions; this choice yields the hardest
jet energy spectrum of a number
of $Q^2$ scales considered. Two different techniques are used to
transform the partons produced by VECBOS into hadrons and jets, which can
then be processed through the CDF detector simulation. One method employs
a Field and Feynman fragmentation function ~\cite{8} (``SETPRT''), tuned
 to reproduce the features of observed inclusive QCD jets.
The other (``HERPRT''~\cite{10}) uses part of
the QCD shower evolution Monte Carlo
HERWIG. In this case the events generated by VECBOS are
assigned an appropriate flavour and colour configuration, and are processed
through the HERPRT initial and final state evolution program. Unless otherwise
noted, the results presented here will use the HERWIG approach.
The Monte Carlo events have then been processed through a full
simulation of the CDF detector and reconstructed in the same manner as the
data. \\
\begin{figure}
\mbox{
\centering
\epsfxsize 6 in
\epsfysize 6 in
\epsffile[0 190 624 624]{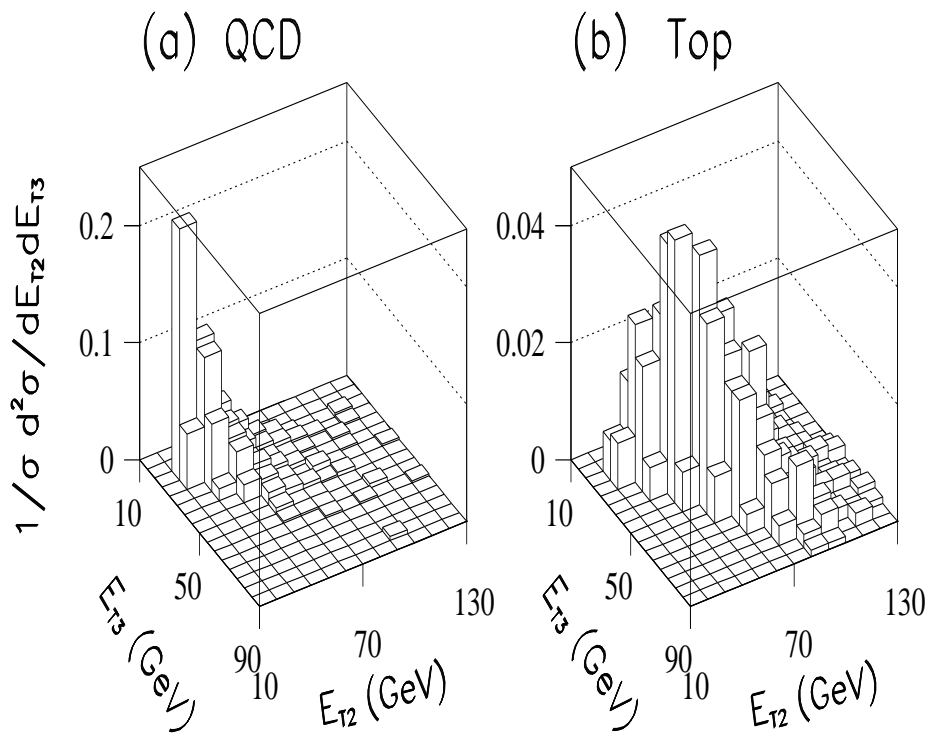}}
\caption{ $\frac{1}{\sigma}$d$^{2}\sigma$ / dE$_{T2}$dE$_{T3}$ for (a)
QCD $W+3$ jet and (b) top quark
(M$_{top}$ = 170 GeV/c$^{2}$) Monte Carlo events.}
\label{et2dim}
\end{figure}
\indent The choice of suitable kinematic parameters to distinguish top quark
events from background is presently the subject of considerable
work.
The D0 collaboration has used event aplanarity and the scalar
 sum of the jet transverse energies~\cite{D0}.
CDF has studied these variables as well as other parameters
 involving combinations of jet energies and angles~\cite{kondo}.
Work is presently in progress to identify which
parameters provide optimal information. In this study we have focused
the analysis on jet transverse energies and polar angles. Our studies
have indicated that these variables are among the most powerful at
separating top quark signal from direct $W +$ jet background. \\
\indent Jets are ordered in $E_T$ with jet$_1$ having the highest
energy, $E_{T1}$.
 It was found that the $E_{T2}$ or $E_{T3}$ variables
 are better discriminant between QCD background and top quark events
than  $E_{T1}$.
 A qualitative indication of the separation that can be
obtained between $t\bar t$ and direct $W +$ jet production on the basis of
$E_{T2}$, $E_{T3}$ is shown in Figure 1, which presents the predicted
density $\frac{1}{\sigma}$d$^{2}$$\sigma$/d$E_{T2}$d$E_{T3}$ of
$W +$ $\ge$ 3 jet and $t\bar t$
events (M$_{top}$ = 170 GeV/c$^{2}$). The distributions are different for heavy
top quark and background events, with $t\bar t$ events characterized by higher
$E_{T}$ jets. \\
\indent Selection of events based on the presence and energy of a fourth
jet is also predicted to be a good discriminant between $t\bar t$
and direct $W$ production.
However, in this analysis we do not {\it require} a fourth jet. This is
done in order to minimize: (a) uncertainties in the theoretical calculation
 of the E$_{T4}$ spectrum in $t \bar t$ events with accompanying
gluon radiation, (b) the uncertainty in the reconstruction efficiency and in
 the energy measurement of low energy jets. The presence of a fourth jet
 will later be examined in this paper as an indication of whether top quark is
present.\\
\indent Another variable which can discriminate between $W+$ jets
and $t\bar{t}$ is $|\cos \theta^{*}|$~\cite{3},
 where $\theta^*$ is the angle between a jet and the
incident
proton direction in the center of mass of the hard subprocess.
The component of the hard subprocess center of mass velocity
 along the beam direction
is calculated using the four--momenta of the W and
all jets with $E_{T}$ $>$ 15 GeV.
 Jets are included down to this relatively low energy in order to
reconstruct the laboratory velocity of the initial state subprocess as well
 as possible.
 In calculating the $W$ 4--momentum,
the longitudinal component of the neutrino cannot be
 determined unambiguously and for simplicity is taken to be zero.
The expected distribution of the jets as a function of
 $|\cos \theta^{*}|_{max}$, the maximum of
$|$cos$\theta$$^{\star}$(jet$_i$)$|$, i=1,2,3
 is shown in Figure~\ref{costheta}(a).
The inclusive jet distribution for direct $W$
events is peaked in the forward direction while that for top quark events is
more central.  As in Ref.~\cite{prd}, jet$_1$, jet$_2$ and jet$_3$
 are required to have   $|$$\eta$(jet)$|$ $<$ 2.
 The  $|\cos \theta^{*}|$ distribution after this cut
 is shown in Figure~\ref{costheta}(b). After the $|$$\eta$(jet)$|$ $<$ 2
 cut,  our studies indicate that a $|\cos \theta^{*}|$ cut still
improves the signal/background ratio. It also allows one to define
a background depleted ``signal sample'' as those
events in which each of the three leading jets satisfies
$|\cos \theta^{*}|<0.7$, and a background enriched ``control sample''
which contains all events
in which at least one of the jets has $|\cos \theta^{*}| > 0.7$.
\begin{figure}
\mbox{
\centering
\epsfxsize 5 in
\epsfysize 5 in
\epsffile[ 0 0 510 510]{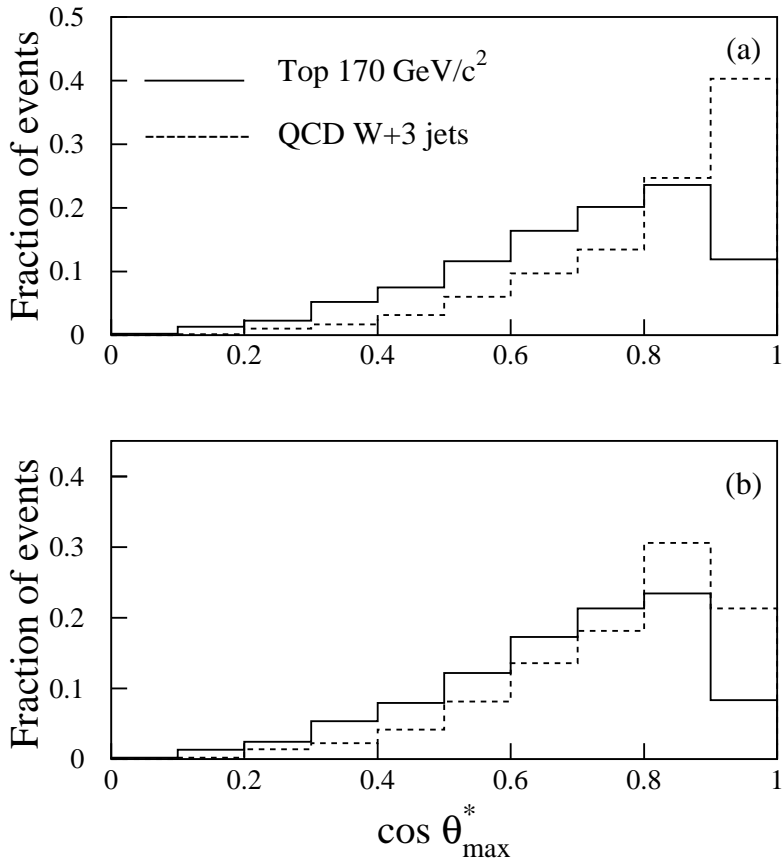}}
\caption{ Distributions of the $|\cos~\theta^{*}|_{max}$ variable predicted
by the HERWIG top quark
(M$_{top}$ = 170 GeV/c$^{2}$) and VECBOS $W+3$ jet calculations for:
(a) the inclusive distribution and
(b) after applying the cut on $|$$\eta$(jets)$|$ $<$ 2.
The distributions are normalized to unit area.}
\label{costheta}
\end{figure}
The Monte Carlo predictions show that the $|$$\cos \theta^{*}$$|$ cut
 generates a  harder jet E$_T$ distribution for top quark production,
while for direct $W +$ jet production it leaves
the E$_T$ distributions essentially unaffected.
Therefore an analysis which attempts to separate top quark from
background based on the shape of the E$_T$ distributions can be
expected to become
 more discriminating after applying the $|$$\cos \theta^{*}$$|$ cut.\\
\indent Using the Standard Model $t \bar{t}$ cross section from
 Ref.~\cite{laenen} we expect approximately 6 events in the signal sample
 and 7 in the control sample for a top quark mass of 150 GeV/c$^2$, while for
a top quark mass of 190 GeV/c$^2$ we expect approximately two events each in
 the signal and control samples.
Assuming the top quark production cross section
from Ref.~\cite{prd}, the number of top quark
events expected for M$_{top}$ = 170 GeV/c$^2$
is 7.7$^{+3.5}_{-2.8}$ in the signal sample and
 9.0$^{+4.1}_{-3.2}$ in the control sample.
 The signal sample contains 14 events and the control sample 35 events.
Therefore in the signal sample a signal/background of the order of 1 can be
expected,
while in the control sample this ratio would be nearly three times worse.

\section{Reliability of $W+$ Multijet Predictions}

\hspace{0.5cm} In subsequent sections a detailed comparison is made
 for the observed jet
energy distributions for events that contain a $W$ and $ \geq 3$ jets with the
predictions of QCD direct production of $W$ and jets
(as implemented in the VECBOS program). It is therefore
important to investigate to what extent these predictions are reliable.
 Previously CDF has compared the cross section for $W+n$ jet production
 ($n \leq$ 4)
 with QCD predictions~\cite{wjetprl} and found good agreement.
In addition, the jet energy distributions and
rapidities for $W +1$ jet and $W+2$ jets
 show  good agreement with the QCD calculations~\cite{wjetprl}.
The UA2
collaboration at CERN has examined the transverse energy distributions
for multijet events with up to six final state partons and found good
agreement ~\cite{ua2jets} with expectations from QCD.
 CDF has also found excellent agreement between
observation and QCD predictions for inclusive distributions in
3 and 4 jet data samples~\cite{3jetprd}\cite{4jetprd}. Although
 the UA2 comparison and the CDF multijet comparison
 involve a different set of matrix elements than for
 jets associated to $W$ production,
they demonstrate that in terms of jet detection and reconstruction,
 excellent agreement is obtained between observations and theory for events
containing as many as four jets.\\
\indent A good test is provided by the CDF $W+\geq2$ jet data sample, which
 has relatively high statistics, is kinematically  similar to the
$W+\geq3$ jet sample, and has a relatively small fractional contribution
from top quark.
\begin{figure}
\centering
\epsfxsize 5 in
\epsfysize 5 in
\epsffile[ 0 0 510 610]{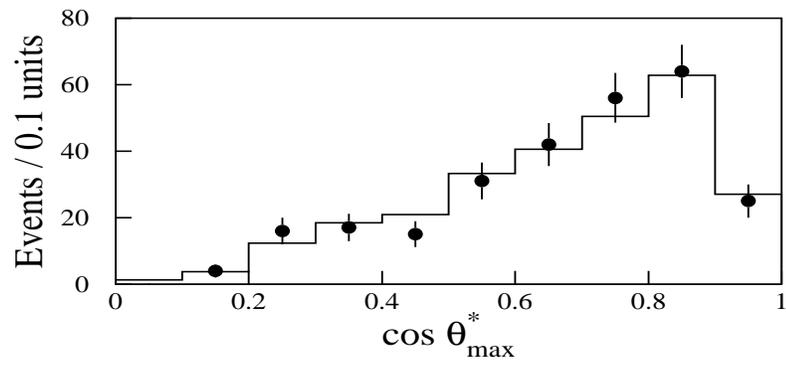}
\caption{ $|\cos~\theta^{*}|_{max}$ for $W + \geq$ 2 jets showing data
(points with error bars) and
 VECBOS $W + 2$ jet events (histogram).}
\label{cos}
\end{figure}

\begin{figure}
\mbox{
\centering
\epsfxsize 5 in
\epsfysize 5 in
\epsffile[ 0 0 510 510]{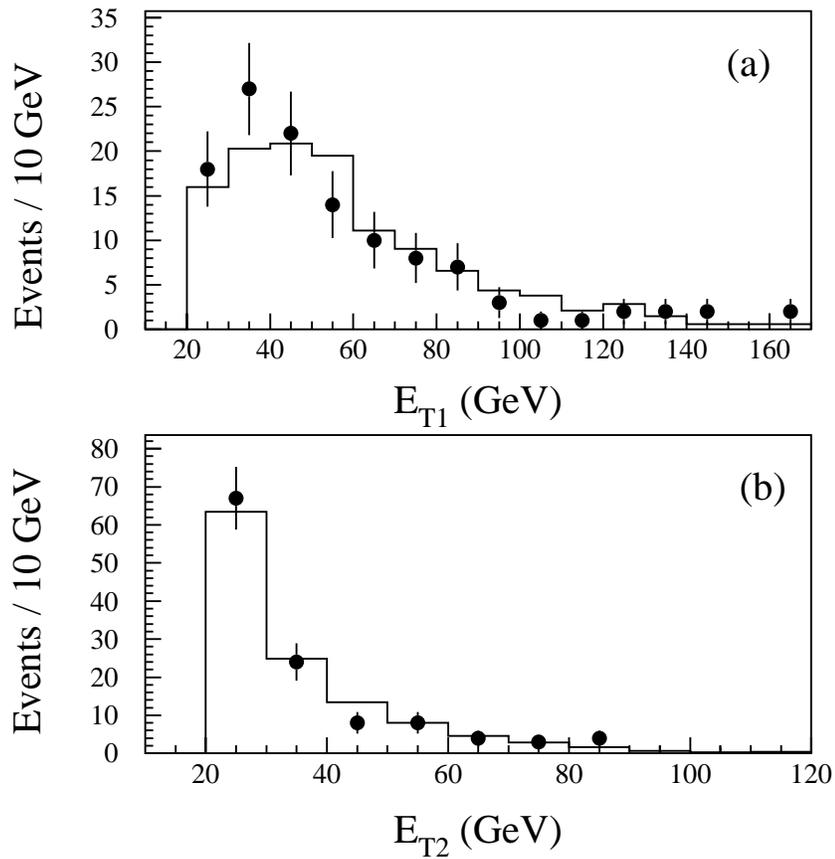}}
\caption{ E$_T$ distributions of $W+\geq 2$ jet showing data (points with
error bars) and VECBOS $W+2$ jet events (histograms). (a) leading jet,
(b) second to leading jet.}
\label{w2jet}
\end{figure}

\begin{figure}
\mbox{
\centering
\epsfxsize 5 in
\epsfysize 5 in
\epsffile[ 0 220 425 567]{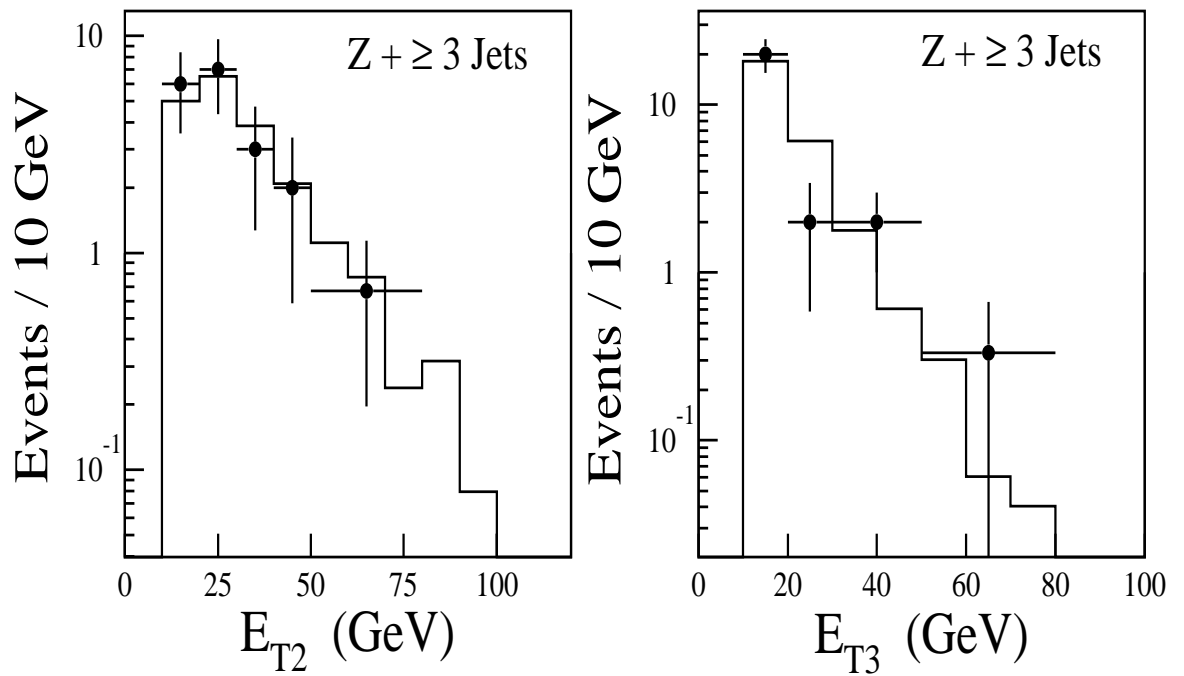}}
\caption{ E$_T$ distributions of second and third leading jets
 in $Z+\geq 3$ jet (inclusive) data (shown as points with error bars)
and VECBOS $Z+3$ jet events (histograms).}
\label{z0jet}
\end{figure}

The angular distributions of data and VECBOS events
are compared in Figure~\ref{cos} in terms of the variable
$|\cos \theta^{*}|_{max}$, which is here the maximum of
$|$cos$\theta$$^{\star}$(jet$_i$)$|$, i=1,2.
As for the $ W + \geq 3$ jet sample, a cut on the jet rapidity was applied
at  $|$$\eta$(jet)$|$ $<$ 2.
The Monte Carlo prediction is normalized to the data. The agreement is
excellent.
The E$_{T1}$ and E$_{T2}$ distributions of these $W + \geq 2$ jet events
 are compared to VECBOS predictions
in Figure~\ref{w2jet} under the requirement that
both jets
have $|$$cos \theta^{*}$$|$ $<$ 0.7. The Monte Carlo prediction is normalized
 to the data.
The confidence level of the likelihood that the data are
consistent with Monte Carlo predictions is 55\% for Fig. 4(a) and 69\% for
Fig. 4(b).
While the VECBOS calculation has no phenomenological parameters,
 the results do depend on the
choice of the $Q^2$ scale and on the  minimum
separation and $P_T$ of the generated jets.
 Use of $Q^2 = < P_T>^2$ instead of $Q^2 = M_W^2$ yields  softer spectra.
 The agreement between data and predictions
 is equally good. With the choice   $Q^2 = < P_T>^2$
 the confidence level of the likelihood that the data are
consistent with Monte Carlo predictions is 62\% for both the E$_{T1}$
and E$_{T2}$ distributions.
 Since top quark events are expected to give harder jet energy spectra
than direct W+jet production, our default choice,
 $Q^2 = M_W^2$, (see Section 3) is conservative.
Additional tests on the sensitivity to the $Q^2$ scale
of $W+\geq 3$ jet Monte Carlo predictions are
shown in the next section. Finally, a test of the
predictions of jet production associated with
 vector bosons  may be obtained from the $Z+ \geq 3$ jets
sample, where little contamination from Standard Model
top quark events is expected.
Data and predictions are shown in
Figure~\ref{z0jet}.
While the statistics are limited, the agreement is good.
In the higher
statistics $Z+ \geq 2$ jet sample, one also finds good agreement between
data and predictions for the E$_T$ distributions of the two leading jets.

\section{Kinematic Analysis}

\begin{figure}
\mbox{
\centering
\epsfxsize 5 in
\epsfysize 5 in
\epsffile[ 0 0 482 624]{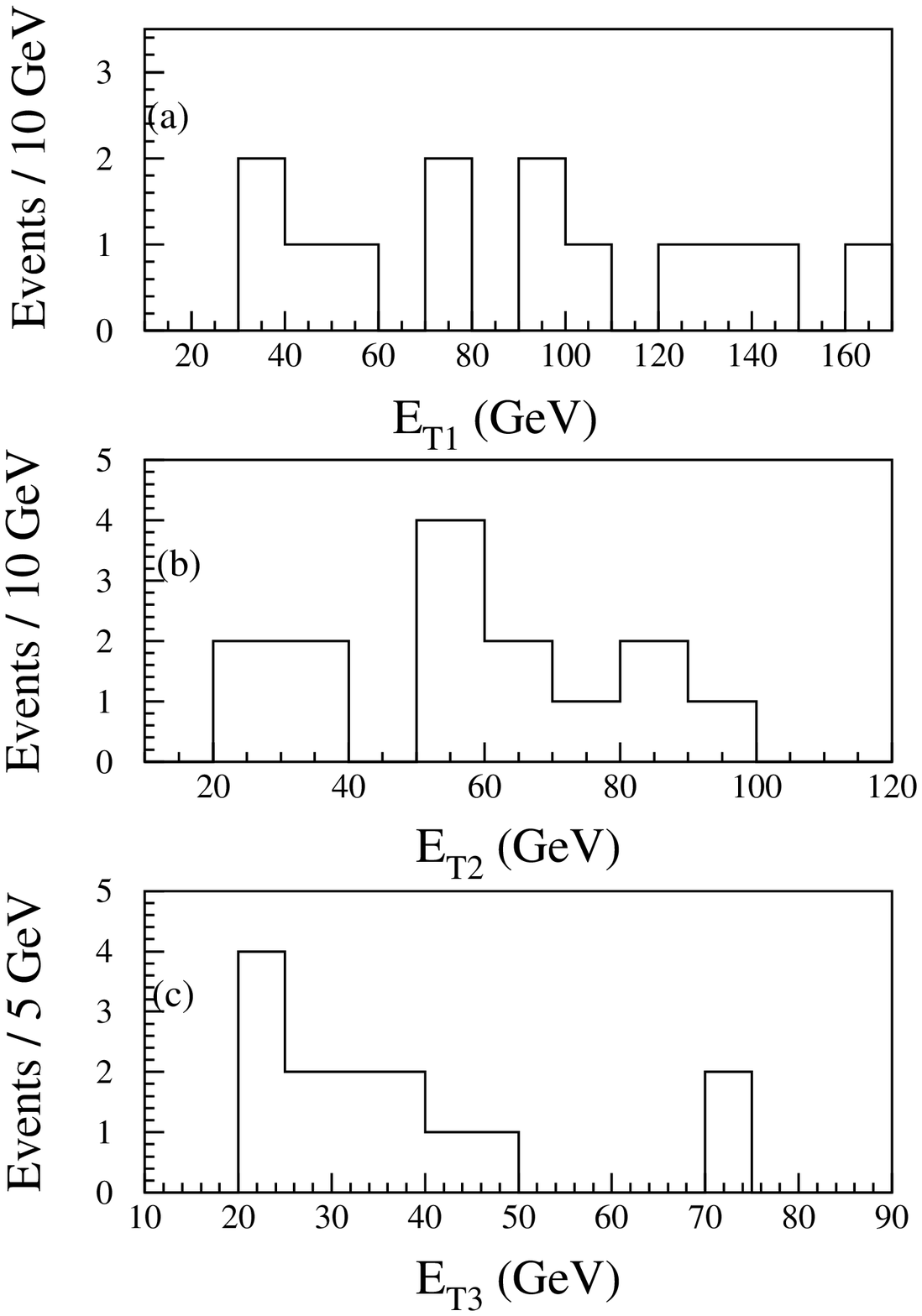}}
\caption{ Jet energy distributions for the three leading jets in the
 14 events passing the signal sample selection cuts.
There is one overflow in E$_{T1}$ at E$_{T1}$ = 224 GeV.}
\label{etdata}
\end{figure}

\begin{figure}
\mbox{
\centering
\epsfxsize 5 in
\epsfysize 5 in
\epsffile[ 0 0 482 624]{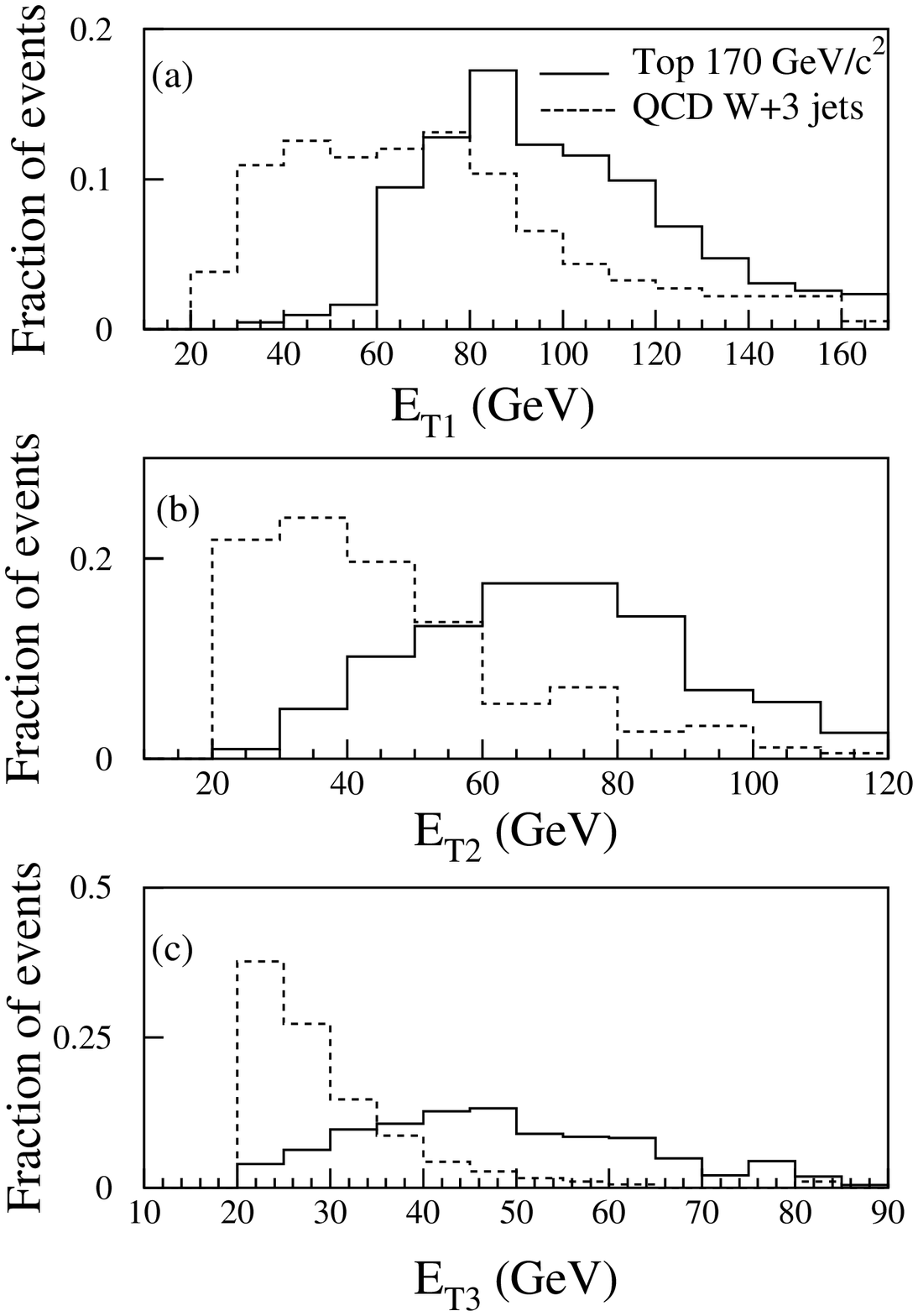}}
\caption{ Jet energy distributions for HERWIG top quark (solid line)
and VECBOS $W+3$ jet events
(dashed line) passing the signal sample selection cuts. Each distribution
is normalized to unit area.}
\label{etmont}
\end{figure}

\hspace{0.5cm} As discussed in Section 3, a signal sample of 14 events
 is defined by the requirement that
the three leading jets have  $|$cos$\theta$$^*$(jet)$|$ $<$ 0.7.
The background enriched
control sample, where at least one jet has $|$cos$\theta$$^*$(jet)$|$ $>$
0.7 contains 35 events.
 Figure~\ref{etdata} shows the E$_T$ distributions of the three leading
jets in the signal enriched sample.
 Figure~\ref{etmont}
 shows the same distributions for
the process $W+3$ jets as predicted by VECBOS and for top quark production
modeled with the HERWIG Monte Carlo at a top quark mass of 170 GeV/c$^2$. The
distributions are normalized to unit area.
The E$_T$
distributions of the data are harder than those expected from VECBOS.
 To combine the information from both E$_{T2}$ and  E$_{T3}$,
 a discriminating function, ``absolute likelihood'', is defined as follows:
\begin{equation}
 L_{abs} = (\frac{1}{\sigma} \frac{d\sigma}{dE_{T2}}) \times
(\frac{1}{\sigma} \frac{d\sigma} {dE_{T3}})
\end{equation}
that is,
as the product of the two differential transverse energy distributions each
normalized to unit area.
 The  $\frac{d\sigma}{dE_{T}}$
are derived from the Monte Carlo simulated distributions fitted by
analytical functions.
A L$_{abs}$ function can be defined for any process for
which a model exists, in particular for QCD $W+\geq 3$ jets (L$_{abs}^{QCD}$)
and top quark (L$_{abs}^{top}$).
 The L$_{abs}$'s can be combined to define a ``relative likelihood''
(L$_{rel}$) for top quark
 versus QCD as
\begin{equation}
L_{rel} = L_{abs}^{top}/L_{abs}^{QCD}.
\end{equation}
 Note that the absolute likelihoods are not probabilities, since E$_{T2}$ and
 E$_{T3}$ are correlated.
The relative likelihood allows one to compare each
individual event to the expectation from QCD and from top quark
in terms of a single number.
This ``kinematic tag''  provides a natural definition of the
 cut which discriminates events which are more top quark--like from
events which  are more QCD--like.
 A possible disadvantage of L$_{rel}$ is its dependence on M$_{top}$.
In particular, a L$_{rel}$ which is optimized for a certain top quark
mass may have reduced
sensitivity if the actual top quark mass is significantly different
from the assumed mass. We choose for our analysis
M$_{top}$ = 170 GeV/c$^{2}$, based on the results of Ref.~\cite{prd}.
 We discuss the effect of a possible different choice in
 Section 5.1.
\subsection{Data--Monte Carlo comparison}

\begin{figure}
\epsfxsize 5 in
\epsfysize 5 in
\mbox{
\centering
\epsffile[0 0 482 624]{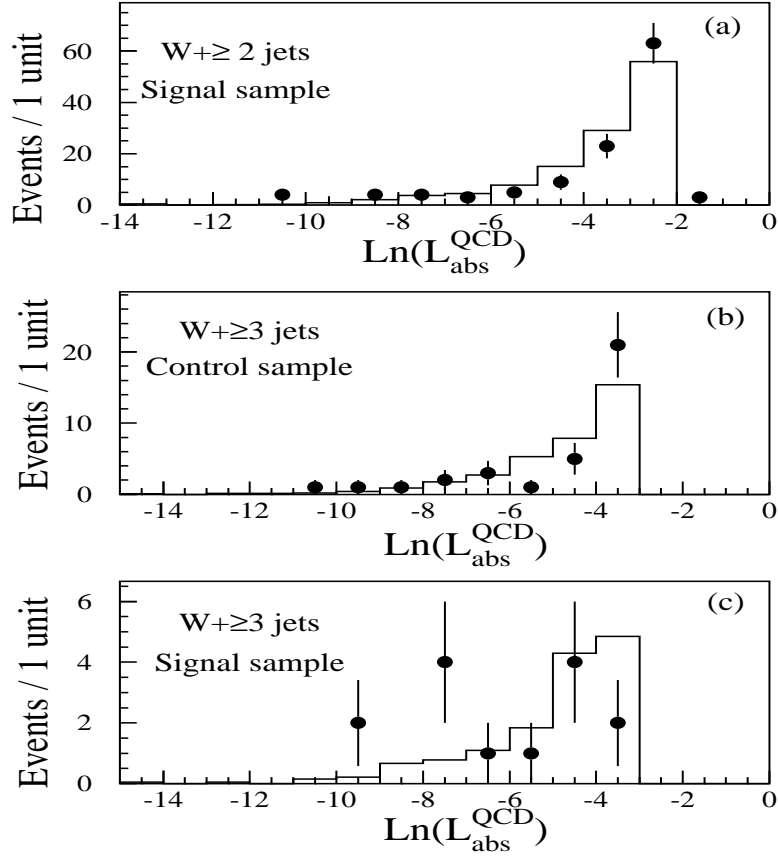}}
\caption{ QCD-predicted absolute likelihood distributions (histograms)
compared to data (points with error bars). (a): the two leading jets in
$W+\geq 2$ jets, a cut $|$cos$\theta$$^{*}$$|$ $<$ 0.7 was applied to jet$_1$
and jet$_2$;
 (b): jet$_2$, jet$_3$ in the control sample; (c) : jet$_2$, jet$_3$
in the signal sample.}
\label{al}
\end{figure}

\begin{figure}
\mbox{
\centering
\epsfxsize 14.cm
\epsfysize 14.cm
\epsffile[0 0 567 567]{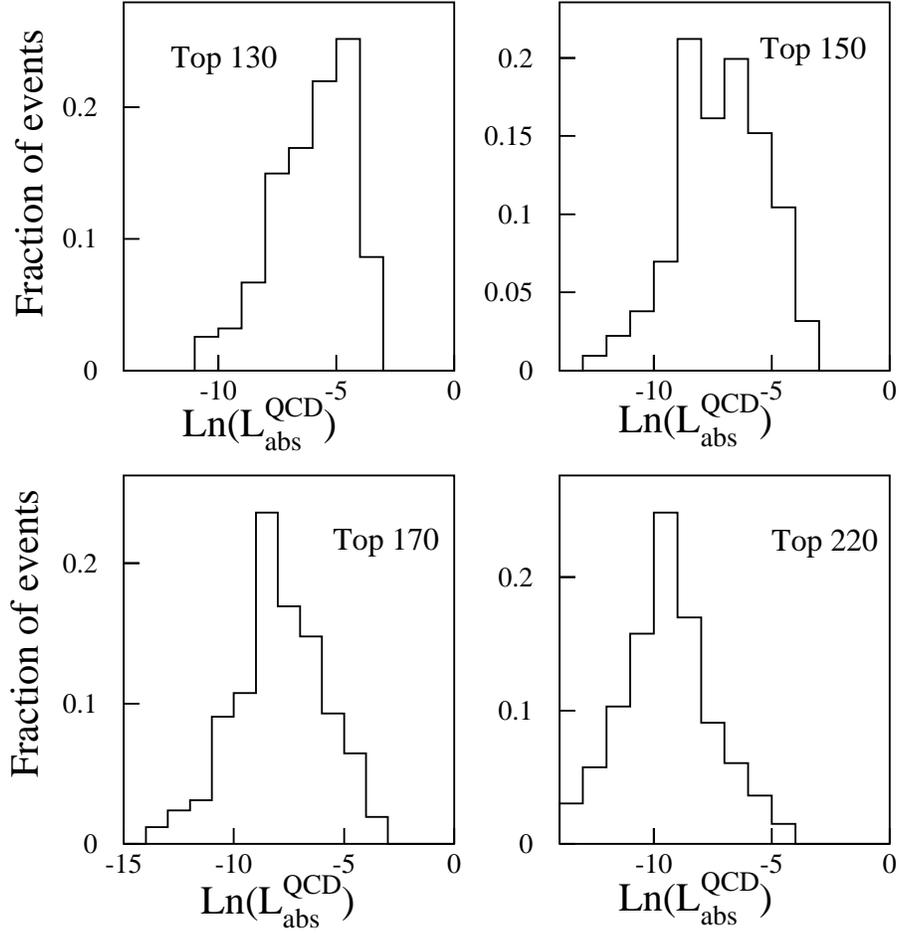}}
\caption{ Expected distributions of $t\bar t$ Monte Carlo (HERWIG)
events as a function
of ln(L$_{abs}^{QCD}$), when interpreted as direct $W +$ jet events.
The distributions are normalized to unit area.}
\label{aLISAJET}
\end{figure}

\begin{figure} \epsfxsize 13.5cm \mbox{ \centering
\epsffile[0 0 510 510]{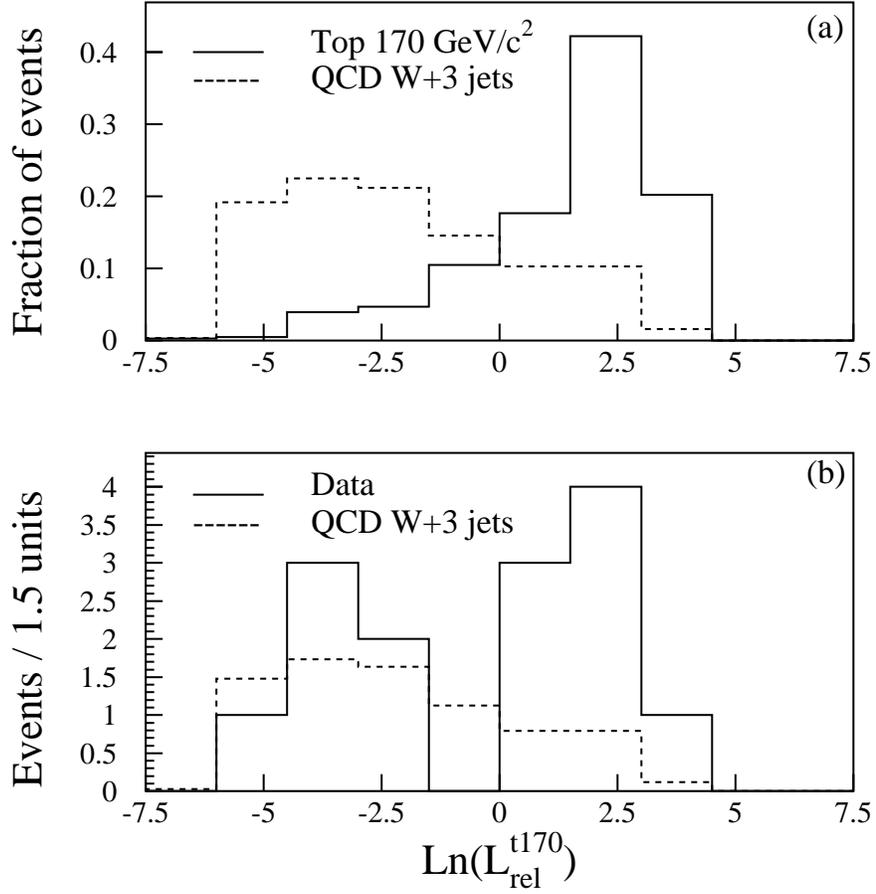}}
\caption{ Ln(L$_{rel}^{t170}$) for QCD VECBOS, top quark
HERWIG (M$_{top}$ =  170 GeV/c$^{2}$) and data events for $W+3$ or
more central jet events (signal enriched
sample). (a): $W+\geq 3$ jet VECBOS (dotted histogram) and top quark
HERWIG  (solid histogram),  normalized to unit area. (b): data (
solid histogram) and VECBOS (dotted
histogram). VECBOS is  normalized to data in the region ln(L$_{rel}^{t170}$)
$<$ 0.}\label{rl_ss} \end{figure}

\begin{figure}
\epsfxsize 13.5 cm
\mbox{
\centering \epsffile[0 0 510 510]{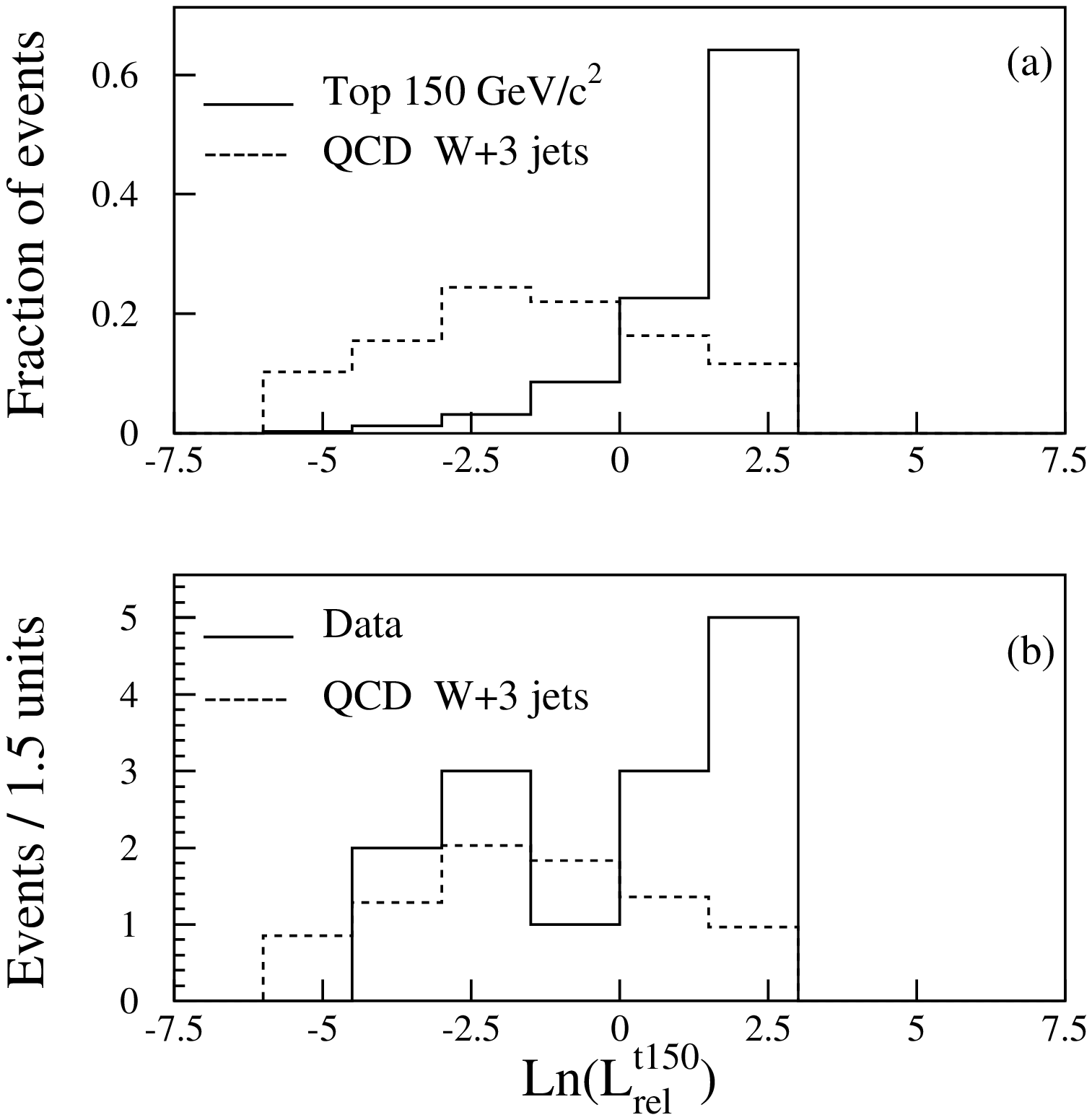}} \caption{ (a)
Expected distribution in ln(L$_{rel}^{t150}$) of QCD and $t\bar t$ events, for
M$_{top}$ = 150 GeV/c$^2$; (b) the data displayed as a function of the same
variable. The dotted histogram shows the QCD distribution
normalized to the data at ln(L$_{rel}^{t150}$)$<$0.}
\label{rl140}
\end{figure}

\hspace{0.5cm} Figure 8 shows a comparison of the expected and observed
distributions for ln(L$_{abs}^{QCD}$)
 for the $W+ \geq 2$ jet and $W+ \geq$ 3 jet signal and control samples.
 In the case of $W+ \geq 2$ jets, the L$_{abs}$ is defined
as the product of the E$_{T1}$  and E$_{T2}$ distributions,
since a third jet is not always present in the event. The $W+ \geq 2$ jet
sample is expected to have a small top quark fraction.
The comparison with the VECBOS prediction (Fig. 8(a)) shows good
agreement. The $W + \geq$ 3 jet control sample data, where the QCD
background is expected
to dominate, also agree with the QCD prediction
as shown in Figure~\ref{al}(b).
In the $W + \geq$ 3 jet signal sample (Figure~\ref{al}(c)), where
a $t\bar t$ contribution could be present,
VECBOS predictions and data are somewhat different.
 In order to check how significant this difference is we performed a
 likelihood calculation, by assuming a Poisson distribution with mean
 equal to the Monte Carlo prediction for each bin.
A confidence level
of 19.6\% is found for VECBOS to agree with the data of Figure ~\ref{al}(b),
and 2.7\% for Figure ~\ref{al}(c). The likelihood for Figure ~\ref{al}(c)
is small enough to suggest
 a component in the signal sample not well described
by VECBOS.  The
expectations for HERWIG $t\bar t$ events, when interpreted as direct $W$+
jet events, are  shown in Figure~\ref{aLISAJET}
for a number of different top quark masses.
 The signal sample data distribution in ln(L$_{abs}^{QCD}$) seen in Figure
{}~\ref{al}(c) is  consistent with a combination of direct $W +$ jet
events and of
$t\bar t$ events
 in a wide mass range around M$_{top}$ = 170
GeV/c$^{2}$.
 Figure~\ref{rl_ss}(a) shows how the VECBOS $W+3$ jet events and
HERWIG top quark events are distributed in relative likelihood,
ln(L$_{rel}^{t170}$), when the cuts of the signal sample are
applied. The symbol L$_{rel}^{t170}$ indicates that M$_{top}$ = 170
GeV/c$^{2}$ was used to predict the expected E$_{T2}$ and E$_{T3}$
distributions.
 The two distributions are separated well enough to make a top quark
signal visible, provided the signal/background is of order 1 as argued in
 Section 3.
Figure~\ref{rl_ss}(b) shows how the data events of the signal
sample are distributed in
ln(L$_{rel}^{t170}$), along with the
VECBOS distribution from Figure~\ref{rl_ss}(a),
normalized to the data at ln(L$_{rel}^{t170}$) $<$ 0.
The data are not distributed as expected from a
pure QCD $W+$jet sample, and look like a superposition of $t \bar t$ and
QCD events.\\
\indent We find
6 events with ln(L$_{rel}^{t170}$) $<$ 0 (more QCD--like) and 8 events with
ln(L$_{rel}^{t170}$) $>$ 0 (more top quark--like). If we normalize the
VECBOS distribution,
78\% of which is expected to have ln(L$_{rel}^{t170}$) $<$ 0, to the 6
events observed in that region, then we expect 1.68$^{+0.85}_{-0.62}$
  VECBOS events
with ln(L$_{rel}^{t170}$) $>$ 0
compared to 8 events observed.
 This excess represents kinematic evidence for the presence of $t\bar t$
production.\\
\indent  Although M$_{top}$ = 170 GeV/c$^2$ was assumed, the result is not
sensitive to the precise value of M$_{top}$.
 As an example,
 Figure~\ref{rl140}(a) shows the expected distributions in
ln(L$_{rel}^{t150}$) for
VECBOS $W + 3$ jets and HERWIG
$t\bar t$ events if M$_{top}$ = 150 GeV/c$^2$
is assumed.
The two samples are still well separated.
The data are distributed as shown in Figure
{}~\ref{rl140}(b), and still indicate a superposition of the two processes.
The results of a two--component fit to the ln(L$_{rel}^{t170}$)
distribution as a function of M$_{top}$ and the significance of the observed
excess at positive ln(L$_{rel}^{t170}$) will be examined in a later section. \\


\subsection{Evaluation of the statistical significance}
\begin{figure}
\epsfxsize 13.5cm
\mbox{
\centering
\epsffile[0 0 510 510]{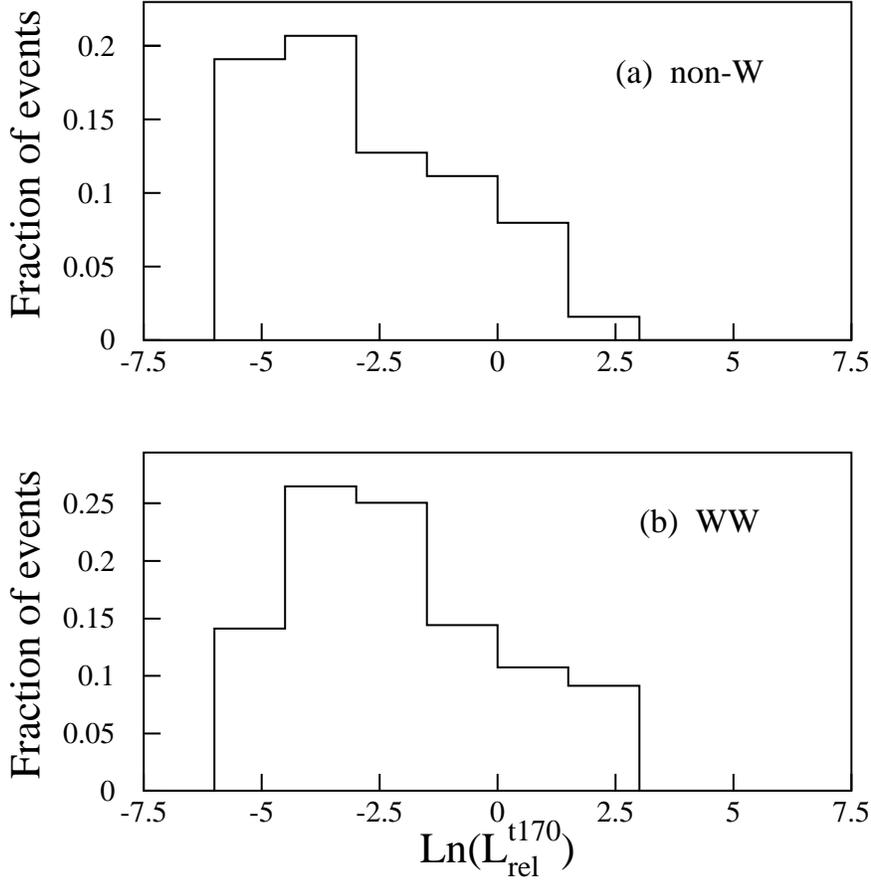}}
\caption{ (a) Ln(L$_{rel}^{t170}$) distribution of non--$W$ background,
 as derived from
the study of a sample of non isolated leptons, with small $\met$;
(b) Monte Carlo predicted
 ln(L$_{rel}^{t170}$) distribution of $WW$ events.
The distributions are normalized to unit area.}
\label{nonwback}
\end{figure}

\begin{figure}
\epsfxsize 13.5 cm
\mbox{
\centering
\epsffile[0 0 510 510]{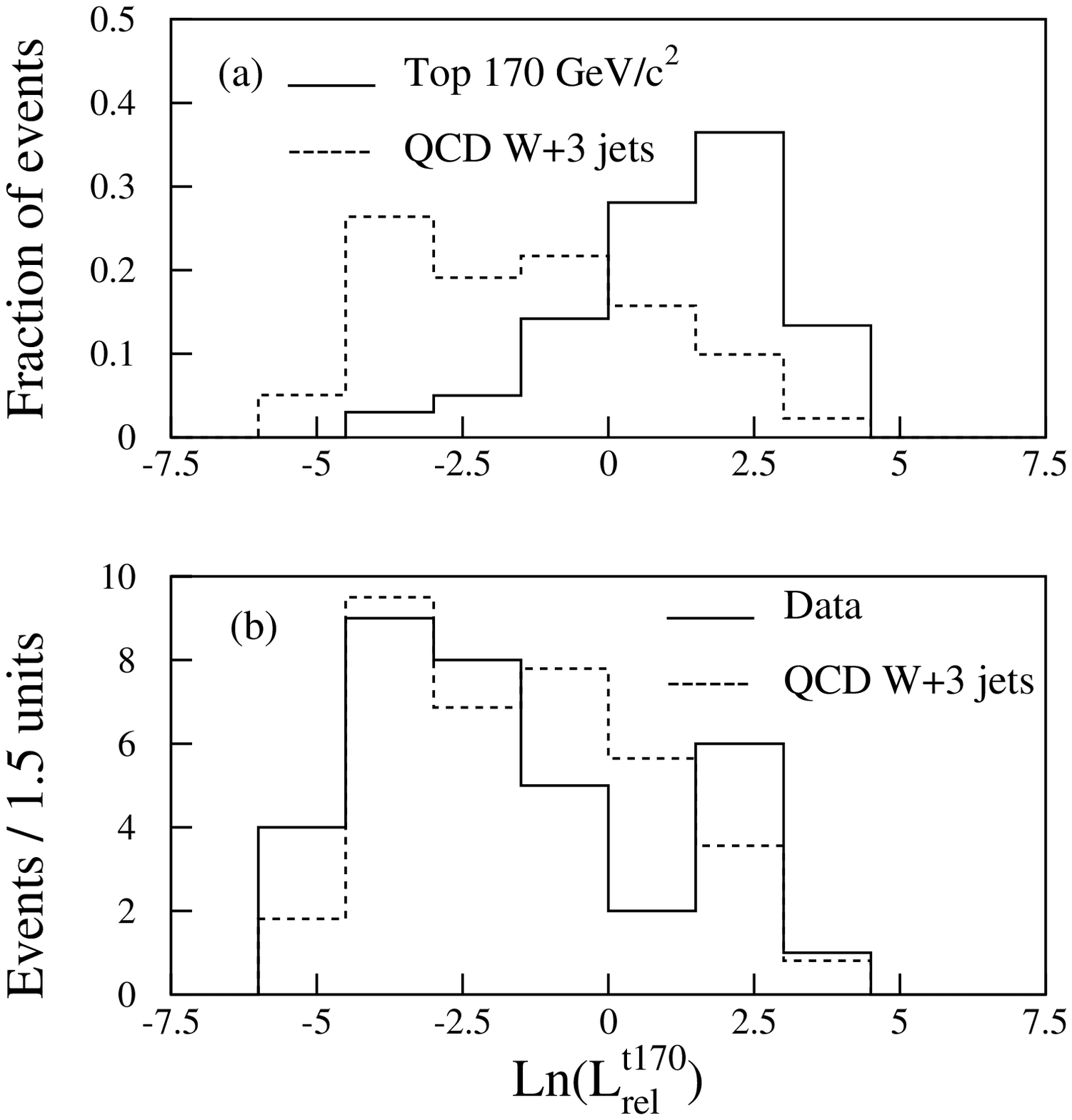}}
\caption{ Distributions of ln(L$_{rel}^{t170}$) for the control sample.
(a) Distributions of VECBOS $W+3$ jet (dotted histogram)
and HERWIG top quark events (solid histogram), normalized to unit area.
(b) 35 data events (solid histogram) versus VECBOS (dotted, with statistical
errors). VECBOS has been normalized to data in the region ln(L$_{rel}^{t170}$)
 $<$ 0.}
\label{rl_cs}
\end{figure}

\hspace{0.5cm} The probability that a background fluctuation can produce
the observed excess at ln(L$_{rel}^{t170}$) $>$ 0 is calculated from the
 binomial probability that given the 14 signal
sample events, they are distributed with 8 or more events in
the positive ln(L$_{rel}^{t170}$) side. The calculation makes use of the
fraction of
 QCD $W +$ jet events expected at ln(L$_{rel}^{t170}$) $<$ 0 and of the
statistical error on the fraction.
For the primary result (shown in line~1 of Table~1)
VECBOS $W+3$ parton production with a choice of  $Q^{2}$ = M$^{2}_{W}$ and
the HERPRT fragmentation is used.
In this case the probability to observe 8 or more events from a
background fluctuation is 0.5\%.
This disagreement between observation and the VECBOS prediction is
large enough to suggest the possibility  that either VECBOS is wrong, or that
there is an additional high E$_{T}$ process present in the data sample. \\
 \begin{table}[htb] \centering \begin{tabular}{|l|c|}  \hline
                         &  \% rate at ln(L$_{rel}^{t170}$)$<$ 0
\\ \hline \hline
 1.~~~~ HERPRT ($Q^2$=M$_W^2$) & 78.2$\pm$2.4 \\ \hline
 2.~~~~ E$_{T}$ scaled down    & 81.2$\pm$2.6  \\ \hline
 3.~~~~ E$_{T}$ scaled up      & 74.9$\pm$2.2   \\  \hline\hline
 4.~~~~ HERPRT ($Q^2$=$<$P$_T$$>^2$)
                         & 80.4$\pm$5.5  \\ \hline
 5.~~~~ SETPRT ($Q^2$=$<$P$_T$$>^2$)
                         & 81.1$\pm$2.8 \\ \hline \hline
 6.~~~~ HERPRT ($Q^2$=M$_W^2$) &  \\
 ~~~~~~ + Systematics + non-$W$ backgrounds
                         & 78$\pm$5 \\ \hline \hline
\end{tabular} \label{tab:prob}
\caption{ Fraction of background events at ln(L$_{rel}^{t170}$) $<$ 0
for different predictions of the ln(L$_{rel}^{t170}$) shape.
The predictions compare to 6 data events observed at
ln(L$_{rel}^{t170}$) $<$ 0 and 8 events at ln(L$_{rel}^{t170}$) $>$ 0.}
\end{table}
\indent Next we address systematic uncertainties in the Monte Carlo
predictions.  One systematic uncertainty is the
possible difference in the energy scale of data and
Monte Carlo events.
 The relative uncertainty in the jet energy scale of the calorimeter decreases
 with increasing the jet energy.
These effects have been taken into account by varying Monte Carlo jet
    energies by an uncertainty
    from $\pm$10\% at 8~GeV to $\pm$3\% at 100~GeV to account
    for detector effects, in quadrature with
    a $\pm$10\% uncertainty due to the assignment
    of energies to partons in the presence of gluon radiation, which is the
 dominant uncertainty.
 The second and third lines of Table 1 show the results.
The uncertainty in VECBOS due to the lack of higher order contributions can
be addressed by changing the $Q^2$ scale in $\alpha_{s}$.
 For comparison with the results shown in the first line of Table 1, the
results
for $Q^2$=$<$P$_T$$>^2$ are shown in the fourth line of Table 1.
The fourth and fifth lines compare the results using our default
fragmentation algorithm HERPRT with SETPRT (see Section 3)
and show very little difference. \\
\indent  Contributions to the
event sample from background sources
other than the dominant direct $W+$jet production
were studied to determine if they could
 explain some of the excess at ln(L$_{rel}^{t170}$) $>$ 0.
These additional backgrounds are of two
types. First, in the $W+$jet sample there is a fraction of non--$W$ events
 (e.g.:
hadrons misidentified as electrons or muons, or real leptons from $\bbbar$).
 As in \cite{prd},
the number of such events is estimated by extrapolating
 the number of events which pass the $\met$ cut but have
non--isolated
leptons, to the region in which lepton isolation is required.
When only the isolation cut is released in the signal sample,
no additional event enters.
Following this procedure,
the signal sample is estimated to contain 0.0$_{-0.0}^{+0.9}$ events
 from this source.
The ln(L$_{rel}^{t170}$) distribution of the non--$W$ events
 is shown in Figure~\ref{nonwback}(a) and
is similar to the Monte Carlo predicted distribution of VECBOS events. \\
\indent  A second background is $WW$, $WZ$ events or single $Z$
 events with one non-identified
decay lepton. The ISAJET Monte Carlo is used to simulate
these backgrounds.  The $WW$ and $WZ$ backgrounds are normalized with the
next--to--leading order computations of the cross section from
Ref.~\cite{ww}.
For the $Z$ background the normalization is provided by the measured
CDF $Z \rightarrow ee$ cross section~\cite{Bib_Z_to_ee}.
The estimated number of such events in the
signal sample is 0.9$\pm$0.3 $WW$ events,
0.13$\pm$0.05 $WZ$ and 0.14$^{+0.06}_{-0.04}$ misidentified $Z$'s. The
 ln(L$_{rel}^{t170}$)
distribution of the dominant $WW$ contribution is shown in
Figure~\ref{nonwback}(b). Again, most events are at ln(L$_{rel}^{t170}$)
$<$ 0, as for the
QCD single $W+$jet background.  (21$\pm$9)\% of non--$W$ events
and (20$\pm$4)\% of $WW$ events are expected at ln(L$_{rel}^{t170}$)
$>$ 0, compared with the (21.8$\pm$2.4)\% for the $W +$ jet background
(top line in Table 1). \\
\indent The probability that the observed excess at positive
ln(L$_{rel}^{t170}$) is consistent with
the VECBOS prediction including the effects of non-VECBOS backgrounds and
 the other systematic errors discussed above is computed as follows.
 Non-VECBOS events are chosen from a Poisson distribution
with the means presented above, and are
distributed at positive or negative ln(L$_{rel}^{t170}$) according to the
determined fraction.
The remainder of the  14~events are taken to have the ln(L$_{rel}^{t170}$)$
<$0 fraction predicted by VECBOS, which is taken to be (78 $\pm$ 5)\%.
As can be seen from Table 1 this adequately allows for the variations
due to changes in the energy scale, Q$^{2}$ scale, and the statistical error.
The probability is
calculated via a Monte Carlo program that includes all the uncertainties
mentioned above.  The probability is 0.8\%
to observe $\geq$ 8 events with ln(L$_{rel}^{t170}$) $>$ 0 in a sample of
14~events
originating from direct $W +$ jets and non-$W$ sources.\\
\indent We tested whether the results are stable under reasonable variations in
the event selection requirements for the signal sample.
When we change the requirement on E$_{T1}$ from 20 to 50 GeV, change the cut on
cos$\theta$$^{*}$ from 0.7 to either 0.65 or 0.75 or change the
jet--jet separation cut from $\Delta R$ = 0.7 to 0.6, we get the probabilities
for a statistical fluctuation of
0.5\%, 1.9\%, 0.5\% and 0.7\% respectively.
In the worst case the background fluctuation probability is 1.9\%.
As an
additional test,
 events were selected with the requirement that the uncorrected
missing transverse energy  $\met^{raw}$ $>$ 20 GeV, and no cut on the
transverse  mass of the $W$.  This is the selection used in
Ref.~\cite{prd}. This
sample has about 3 times larger background from fake $W$ events due to
misidentified leptons,
and the neutrino transverse momentum for real $W$ events
is not as well determined.
 However, the acceptance for $W$ and top quark events is $\simeq$25~\%
larger. This results in a signal sample of
19 events; 11 events are
at ln(L$_{rel}^{t170}$) $<$ 0, 8 at ln(L$_{rel}^{t170}$) $>$ 0.
The probability that $\geq$ 8 events
have ln(L$_{rel}^{t170}$) $>$ 0 if the events were entirely QCD background
is 3.8\%.\\
\indent Figure~\ref{rl_cs} shows how the 35 events of the control sample are
distributed in ln(L$_{rel}^{t170}$), together with the predictions from
VECBOS and HERWIG.
 The data are well described by the VECBOS QCD expectation. There is
no statistically significant indication for an excess at ln(L$_{rel}^{t170}$)
 $>$ 0
(whether this is reasonable is addressed in the next section).
 For this comparison, $Q^2$=M$_W^2$ is used as the
scale for $\alpha_{s}$ and structure functions.
The use of $Q^2$=$<$P$_T$$>^2$ would give a softer
spectrum in ln(L$_{rel}^{t170}$) and the number of predicted QCD events at
ln(L$_{rel}^{t170}$) $>$ 0 would fall by 25\%.
\subsection{Cross Section Calculation}

\hspace{0.5cm} We assume here that the excess of high jet E$_T$ events
in the signal sample
 results from $t \bar t$  production and decay. The most probable
numbers of  $t \bar t$ (N$_{top}$) and $W$ (N$_{W}$) events observed in the
signal-enriched and background-enriched samples are estimated using
an unbinned maximum likelihood fit to the observed ln(L$_{rel}^{t170}$)
distribution. For this fit, we use the ln(L$_{rel}^{t170}$) shapes predicted
from the
HERWIG Monte Carlo program for \TTbar\ events (M$_{top}$ = 170~GeV/c$^2$),
and from the VECBOS Monte Carlo program for QCD $W +$ jet events;
these predicted shapes are shown in Figure~\ref{rl_ss}(a).
 The systematic uncertainties on N$_{top}$
 and the \TTbar\ production cross section, $\sigma_{\TTbar}$,
 are estimated as a function of M$_{top}$.
The following effects are considered:
(1) uncertainty in the jet energy
    scale, estimated as described in Section 5.2;
(2) uncertainty in the Q$^{2}$ scale employed by VECBOS to determine
    the jet  E$_{t}$ spectrum in W events\footnote{We address the
    effect of the Q$^{2}$ scale
    on the VECBOS
    ln(L$_{rel}^{t170}$) shape only. Absolute rate predictions do not
    affect the likelihood procedure.}, estimated
    by choosing Q$^{2}$~=~$<$ P$_{T}>^{2}$ rather than Q$^{2}$~=~M$_{W}^{2}$;
(3) Monte Carlo statistics and uncertainty on the lepton detection efficiency;
(4) choice of $t \bar t$ Monte Carlo generator (HERWIG, PYTHIA, ISAJET);
(5) change of the ln(L$_{rel}$) shape due to variations of assumed top quark
 mass in the range 150 $<$ M$_{top}$ $<$ 190 GeV/c$^{2}$;
(6) the inclusion of non-$W$, $WW$, $WZ$, and misidentified $Z$ contributions
    in the likelihood fit;
(7) uncertainty in the fitting procedure; and
(8) uncertainty in data integrated luminosity (this enters only
    in the calculation of $\sigma_{\TTbar}$).
The results for the signal sample
are summarized in Table~\ref{Table_Syst_Summary}
for $\sigma_{\TTbar}$.
\begin{table}
\centering
\begin{tabular}{|l|ccc|} \hline
 & 150~GeV/c$^2$ & 170~GeV/c$^2$ & 190~GeV/c$^2$ \\ \hline \hline
(1) Jet E$_{t}$ scale
 & +21\% --19\% & +22\% --11\% & +23\% --5\% \\
(2) VECBOS Q$^{2}$
 & +10\% --0\% & +10\% --0\% & +10\% --0\% \\
(3) MC stat + lepton $\epsilon$
 & $\pm$10\% & $\pm$9\% & $\pm$9\% \\
(4) $t \bar t$ generator
 & $\pm$10\% & $\pm$9\% & $\pm$10\% \\
(5) non-$W$ background   & $\pm$2\%   & $\pm$2\%   & $\pm$2\%  \\
(6) fitting procedure  & $\pm$3\%   & $\pm$3\%   & $\pm$3\%  \\
(7) Luminosity         & $\pm$3.6\% & $\pm$3.6\% & $\pm$3.6\%  \\ \hline\hline
N$_{top}$ total
 & +27\% --21\% & +28\% --17\% & +24\% --16\% \\
$\sigma_{\TTbar}$ total
 & +31\% --24\% & +33\% --23\% & +29\% --20\% \\ \hline
\end{tabular}
\caption{ Individual systematic errors on $\sigma_{\TTbar}$
    in the signal sample
    as a function of M$_{top}$ are listed in rows (1) to (7).
    Total systematic uncertainties
    for both N$_{top}$ and $\sigma_{\TTbar}$ are summarized in the
    final two rows.}
\label{Table_Syst_Summary}
\end{table}
The systematic uncertainties on  N$_{top}$ are similar to those
on $\sigma_{\TTbar}$; only the totals are listed in
Table~\ref{Table_Syst_Summary}.
The number of $t \bar t$ events is found to be
N$_{top}$=6.4$^{+3.8}_{-3.2}$ $^{+1.8}_{-1.1}$ and
N$_{top}$ = 0.8$^{+5.3}_{-0.8}$ $^{+4.2}_{-0.8}$ for the signal and control
samples respectively, where the first error is statistical and the second
 error is systematic.

\begin{figure}[hpt]
\epsfxsize 13.5 cm
\epsffile[0 0 567 567]{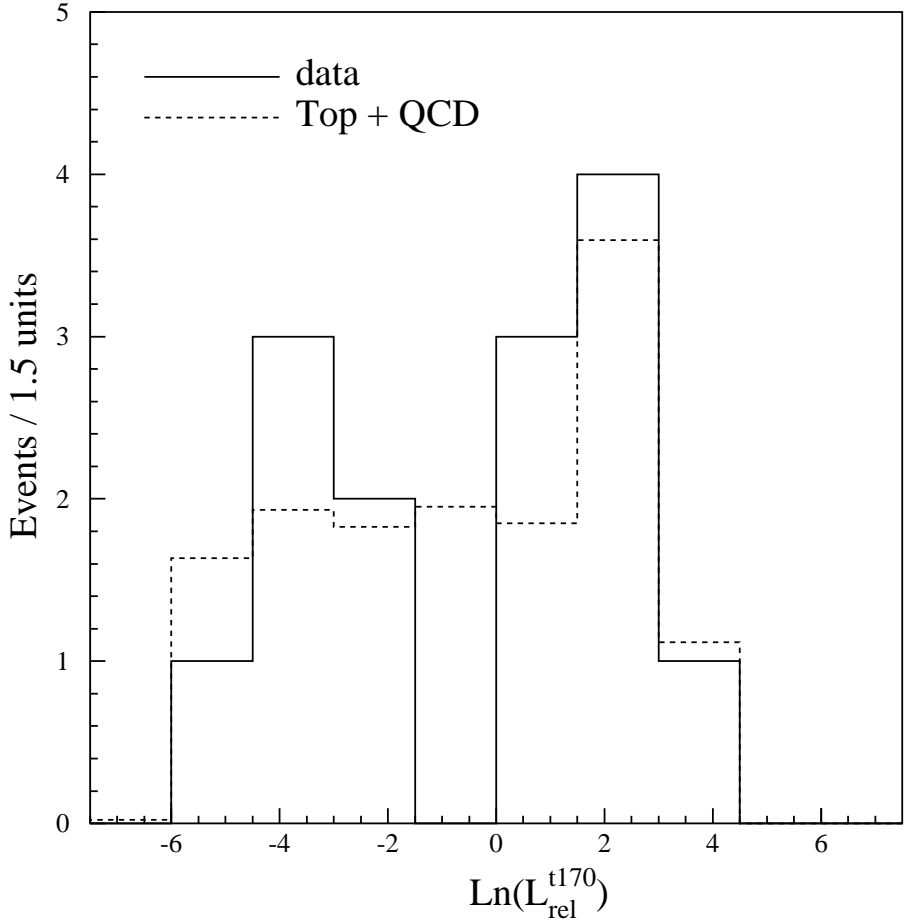}
\caption{ The combined Herwig (M$_{top}$ = 170~GeV/c$^2$) + VECBOS
   ln(L$_{rel}^{t170}$) distribution
   corresponding to the values of N$_{top}$ found
   (for M$_{top}$ = 170~GeV/c$^2$)
   in Table~\protect\ref{Table_Result_Sigtt} (dashed)
   along with the data (solid).}
\label{Fig_Best_Fit_Draft}
\end{figure}

   The fits indicate more $t \bar t$ candidate events
in the signal--enriched sample than in the background--enriched sample.
 The ratio of top quark events in the control sample to top quark
 events in the signal sample  predicted by the Monte Carlo calculation is
1.17
 (9.0/7.7). The data fit finds
 0.13. However, the statistical significance of the
 difference from expectation, taking into account the errors,
 is within 1 $\sigma$.
 The systematic and statistical errors in the
determination of N$_{top}$ are significantly larger in the control sample,
due to the larger number of QCD $W+$jet events.
{}From N$_{top}$, we calculate the
corresponding $t \bar{t}$ cross section.
 This analysis is performed on the signal sample
to minimize the systematic effects
from the uncertainties in
the prediction of the ln(L$_{rel}^{t170}$) shape for QCD $W+$jet events.
 Table~\ref{Table_Result_Sigtt} shows the total
\TTbar\  acceptance, including branching ratios, lepton detection
efficiencies and energy scale uncertainty, as a function of the top quark
mass, and the results for both
 N$_{top}$ and $\sigma_{\TTbar}$.
\begin{table}
\centering
\begin{tabular}{|c|ccc|} \hline
  & M$_{top}$=150~GeV/c$^2$ & M$_{top}$=170~GeV/c$^2$ & M$_{top}$=190~GeV/c$^2$
\\ \hline
               &               &               &            \\
$t \bar t$ acceptance  & 2.7$\pm$0.2\% & 2.9$\pm$0.2\% & 3.0$\pm$0.2\% \\
                &               &               &            \\
N$_{top}$
 & 7.9$^{+4.6}_{-3.8}$ $^{+2.1}_{-1.6}$
 & 6.4$^{+3.8}_{-3.2}$ $^{+1.8}_{-1.1}$
 & 5.6$^{+3.4}_{-2.8}$ $^{+1.4}_{-0.9}$ \\
                &               &               &            \\
$\sigma_{\TTbar}$
 & 15.2$^{+8.7}_{-7.3}$ $^{+4.7}_{-3.7}$ pb
 & 11.6$^{+7.0}_{-5.7}$ $^{+3.8}_{-2.7}$ pb
 &  9.6$^{+5.9}_{-4.8}$ $^{+2.8}_{-1.9}$ pb \\ \hline
\end{tabular}
\caption{ First line: $t \bar t$ acceptance for the signal sample.
 Second and third line: N$_{top}$ and
\TTbar\ production cross section $\sigma_{\TTbar}$
    as a function of M$_{top}$.
    The first error on each entry
    is the data statistical error; the second error on each
    entry is the sum in quadrature of all systematic errors
    listed in Table~\protect\ref{Table_Syst_Summary}. }
\label{Table_Result_Sigtt}
\end{table}
Figure~\ref{Fig_Best_Fit_Draft} shows the summed ln(L$_{rel}^{t170}$)
distribution for $t \bar t$ and VECBOS corresponding to the Table 3
result (with M$_{top}$ = 170~GeV/c$^2$) compared to the observed
ln(L$_{rel}^{t170}$)
distribution.
The cross section  determined from the signal sample
is consistent with that found in~\cite{prd}
and, given the large errors, is not inconsistent with the value of
5.7$^{+1.0}_{-0.6}$ pb predicted by the theory for a top quark mass of
170 GeV/c$^2$~\cite{laenen}.

\section{Identification of $b$ jets}

\hspace{0.5cm} Each top quark event has two $b$ jets. In contrast,
 direct $W+$ jet events  contain
 $b$ jets only at the level of a few percent~\cite{mlm}. In this section we
use two
different methods to identify the $b$ jets in the event (b tagging).
In the first method, the Silicon Vertex Detector (SVX) is used to detect
$B$ hadrons by reconstruction of secondary vertices separated in the plane
transverse to the beam
from the primary interaction vertex as a result of the long $B$ hadron
lifetime.
The algorithm
used to reconstruct secondary vertices, ``jet--vertexing'', and its
performance are discussed extensively in Ref.~\cite{prd}.
 For top quark events, the tagging efficiency (i.e the efficiency
to tag at least one jet in an event as a $b$-jet, including detector
 acceptance)
is expected from Monte Carlo to be
$24\pm5$ \% in the signal sample and $19\pm5$ \% in the control sample.
The efficiency is larger in the signal sample since more events fall within
the fiducial acceptance of the SVX detector.
{}From the number of top quark events derived from the ln(L$_{rel}^{t170}$)
shape
analysis above, $1.5^{+1.0}_{-0.9}$ SVX tags are expected
from top quark  in the signal sample and $0.15^{+1.30}_{-0.15}$	in the control
sample.
The expected number of SVX tags if no top quark were
 present in the sample is computed in the same way as in Ref.~\cite{prd}.
The dominant contribution is from
$Wb\bar{b}$, $Wc\bar{c}$ and mistags. The background estimate
 assumes that all the events are background and uses a tag
 probability for each jet, based on jet E$_T$, $\eta$ and track multiplicity,
 which is derived from a study of a large sample of
inclusive jet data~\cite{prd}. This contribution is found
to be $0.47\pm0.06$ events in the signal sample and $0.83\pm0.11$
events in the control sample. Note that these estimates are
derived directly from the data and do not rely on Monte Carlo
predictions. Adding the other small background contributions
to the tags ($WW$, $WZ$, $Wc$,
$Z \rightarrow \tau \tau$ and non-$W$ events, see~\cite{prd}),
the total expected number of tags assuming that the data contain no top
 quark events
is $0.58^{+0.12}_{-0.09}$ in the signal sample
and $1.1\pm0.2$ in the control sample.
 In the data, 4 events have a SVX tag in the signal sample (3 in common
 with the events selected in Ref.~\cite{prd})
and 1 event is tagged in the control sample. The probability
that the tagging rate in the signal sample is consistent
with the data being only background is about 0.4~\%. On the
other hand, the observed numbers of tags are consistent
with the mixture of top quark and background events expected from the
kinematic analysis. Including
mistags of top quark events and correcting the background
estimates for top quark content in the sample, a total of $1.9\pm1.0$ tags
 is expected in the signal sample and $1.2^{+1.3}_{-0.3}$ in the control
sample.
These numbers are summarized in Table 4, together with the
probability that the observed tag rate is consistent with the
respective expectation.\\
\indent As discussed in Ref.~\cite{prd}, $b$ jets can also be identified by the
presence of an electron or muon from semileptonic B decay.
 For this ``Soft Lepton Tag'' algorithm (SLT),
 the top quark tagging efficiencies are expected to be $19\pm3$ \% in the
signal
sample and $13\pm3$ \% in the control sample.
 The expected number of SLT tags, assuming the data do not contain top quark,
can be computed in a similar way as for the SVX jet--vertexing
tagging algorithm.
 With all background contributions included,
$1.2\pm0.3$ tags are expected in the signal sample and $1.4\pm0.3$ tags
in the control sample.
\begin{table}[htb]
\renewcommand{\arraystretch}{1.25}
\centering
\begin{tabular}{|l|c|c|c|c|c|}  \hline
Sample   & $\begin{array}{c} {\rm Obs.~} \\ {\rm tags} \end{array} $
& $ \begin{array}{c} {\rm Exp.~ tags~} \\ {\rm backg.} \end{array} $
& $ \begin{array}{c}  {\rm Exp.~tags~} \\ {\rm backg.~ +~ t\bar{t}}
\end{array} $
& $ \begin{array}{c} {\rm Prob.} \\ {\rm backg.} \end{array} $
& $ \begin{array}{c} {\rm Prob.} \\ {\rm backg.~ +~ t\bar{t} }
\end{array} $ \\ \hline \hline
Signal & $\begin{array}{c} {\rm 4} \\ {\rm SVX} \end{array} $ &
0.58 $^{+0.12}_{-0.09}$ & 1.9 $\pm$ 1.0 & 0.4\% &16.6\%\\ \hline
Control & $\begin{array}{c} {\rm 1} \\ {\rm SVX} \end{array} $ &
 1.1 $\pm$ 0.2 & 1.2 $^{+1.3}_{-0.3}$ & 66\% &81.7\%  \\ \hline
Signal & $\begin{array}{c} {\rm 4} \\ {\rm SLT} \end{array} $ &
 1.2 $\pm$ 0.3 & 2.2 $\pm$ 0.9 & 4\% &20.7\% \\ \hline
Control & $\begin{array}{c} {\rm 1} \\ {\rm SLT} \end{array} $ &
 1.4 $\pm$ 0.3 & 1.5 $^{+1.0}_{-0.3}$ & 74\% &84.7\%\\ \hline \hline
\end{tabular}
\label{tab:tag_sum}
\caption{Summary of $b$--tagging results in the signal and control
samples. Also shown are the probabilities that the observed rate is
consistent with background only, or with a mixture of top quark + background. }
\end{table}
 In the data, there are 4 SLT tags in the signal sample (3 of them in common
 with Ref~\cite{prd}) and 1 (also in common with Ref.~\cite{prd}) in the
control sample.
 Two of the four events tagged by SLT in the signal sample
 are also tagged by the SVX jet--vertexing algorithm: one of these two
is in common with Ref.~\cite{prd}.
In the signal sample there is again an excess of tags over
the predicted background.\\
\indent The observed b--tags in the signal sample have a low
probability of being
entirely due to a fluctuation in tagging direct $W +$ jet background events.
On the other hand, the hypothesis that the observed events are a mixture of
background and top quark gives a good description of the observed tagging
rates for both SVX and SLT tagging methods.
\begin{figure}
\epsfxsize 13.5 cm
\mbox{
\centering
\epsffile[ 0 0 482 425]{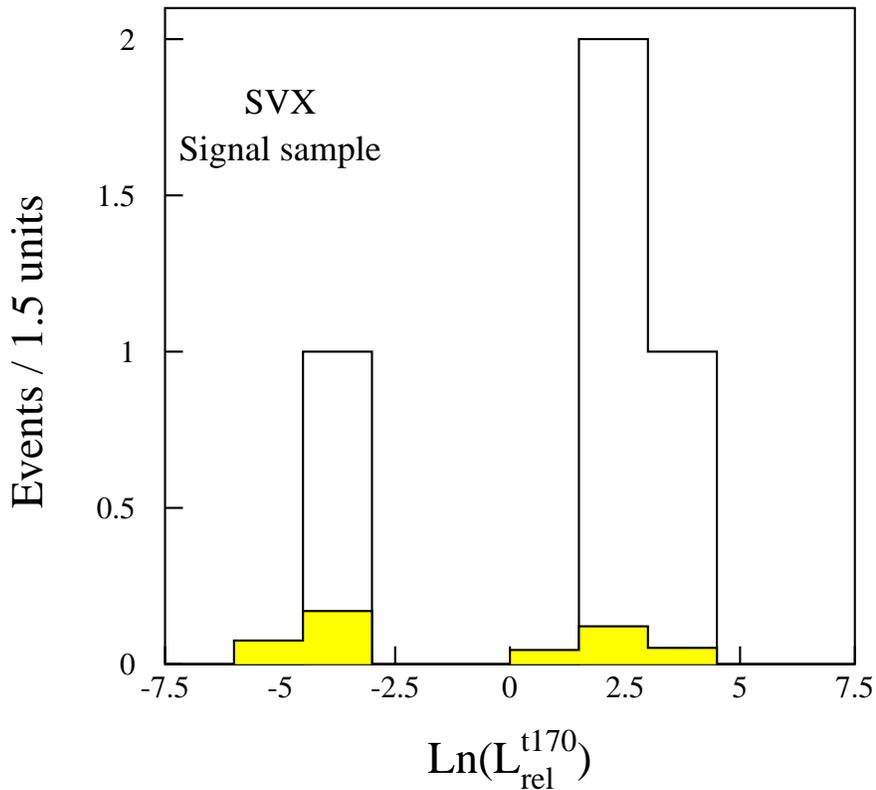}}
\caption{ Distribution in ln(L$_{rel}^{t170}$) of the 4 events of the
signal sample
tagged by the SVX jet--vertexing algorithm.
The expected tags in QCD $W+$jet events
(0.47 in total) are shown as a shaded histogram.}
\label{prd2_hughes}
\end{figure}
\begin{table}[hbt]
\centering
\begin{tabular}{|l|l|cccc|}  \hline
Run--Event    &ln(L$_{rel}^{t170}$)& SVX & SLT &4th jet &prim. lep.\\
\hline\hline
40758--44414 & 3.1&  $\bullet$ &   & *&$e$ \\  \hline
43096--47223 &3.0&$\bullet$ & &*&$e$ \\ \hline
42539--200087&2.2&  $\bullet$ &$\bullet$& *&$\mu$ \\ \hline
43351--266423&1.1& & $\bullet$ &* &$\mu$\\ \hline
45779--6523  &0.7&  &          &*&$e$\\ \hline
42517--44047&-3.5&$\bullet$ &   $\bullet$& &$\mu$\\ \hline
44931--59686 &-4.3&  &              & &$e$ \\ \hline
47616--24577 &-5.0&   &             &&$e$\\ \hline \hline
Run--Event    &ln(L$_{rel}^{t170}$)& SVX&  SLT&4th jet& prim. lep
  \\ \hline\hline
42913--59303  &2.2 &  & & &$e$\\ \hline
45705--54765  &1.6 &  &  $\bullet$ &* &$e$\\ \hline
43276--101844 &0.2 &  & & &$\mu$\\ \hline
45902--240098 &-2.2&  & & & $e$\\ \hline
46290--264893 &-2.9&  & & & $e$ \\ \hline
45801--80320  &-3.4&  & & & $e$ \\ \hline \hline
\end{tabular}
\label{tab:tag_sig}
\caption{ Summary of the signal sample events. The upper section lists
events with at least one jet within the acceptance of the SVX tagging
algorithm. The events in the lower section cannot be tagged by the SVX.
The fifth column identifies those events which have four or more jets,
and the last column identifies the lepton from the W decay as either
an electron or muon.}
\end{table}

\section{Relative Likelihood of b--tagged and four jet Events}

\hspace{0.5cm} The ln(L$_{rel}^{t170}$) values of the signal sample events are
listed in Table 5.
One observes that 5 out of 6 b-tagged events are at ln(L$_{rel}^{t170}$) $>$ 0.
The ln(L$_{rel}^{t170}$) distribution of the 4
 SVX tags together with the dominant
background from $Wb\bar{b}$, $Wc\bar{c}$ and mistags,
 estimated from the inclusive jet
parametrization, is shown in Figure 15.\\
\indent HERWIG predicts that about 80\% of the top quark
events (M$_{top}$ = 170 GeV/c$^2$)
will exhibit a fourth jet in the CDF detector
with transverse energy more than 15 GeV.
Due to the small value of $\alpha_s$, the  fraction of $W + \ge$ 4 jet events
 expected in a $W+ \ge 3$ jet sample is much less.
 Thus the requirement of a fourth jet should further
enrich the sample in top quark events relative to QCD background.
The signal sample contains 6 events with four or more jets
 with E$_T$(jet) $>$ 15 GeV.
 These events are indicated with a * in Table 5. They are all at
 ln(L$_{rel}^{t170}$) $>$ 0.
 Figure 16(a) shows their distribution in ln(L$_{rel}^{t170}$) together
with the prediction from the VECBOS $W+3$ jets +  HERPRT
fragmentation routine.
We recall that in this approach the fourth jet is produced by hard
bremsstrahlung from initial and final state partons.
Studies using $W + 4$ parton + SETPRT Vecbos simulated events
yield similar predictions for background.
\begin{figure}[htb]
\epsfxsize=13.5cm
\begin{center}
\mbox{
\epsffile[0 0 510 510]{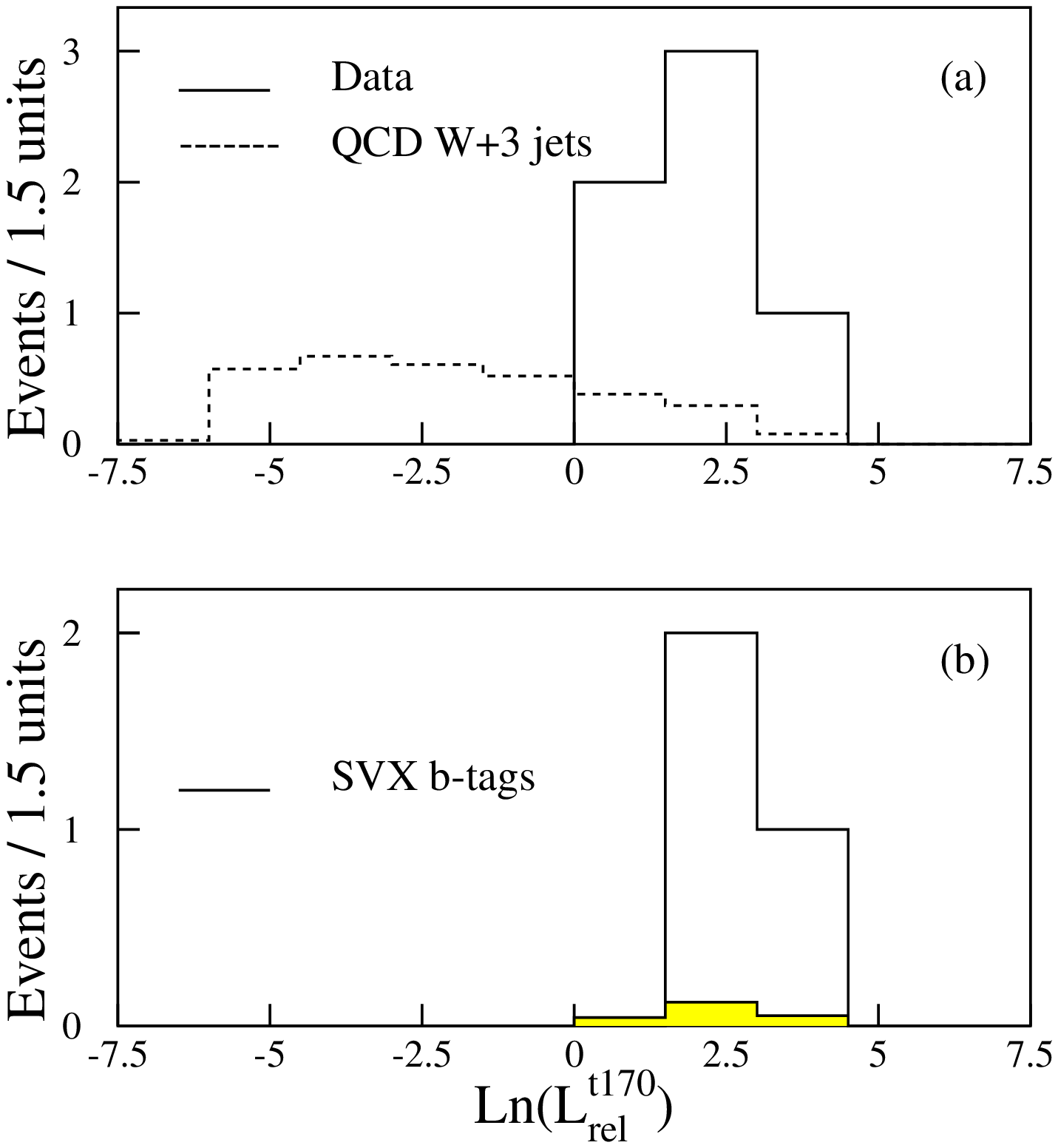}}
\caption{ Distribution in ln(L$_{rel}^{t170}$) of $W+\ge 4$ jet events.
(a) data with VECBOS prediction. (b)
events with a
SVX secondary vertex and prediction (shaded) based on the observed
jets in the events.}
\end{center}
\label{w4j}
\end{figure}
 The normalization chosen for VECBOS in Figure~\ref{rl_ss}(b)
  predicts 2.3 VECBOS events at ln(L$_{rel}^{t170}$) $<$ 0 in Figure 16(a),
while none is observed.
This is  compatible at the 10\% C.L..
 At ln(L$_{rel}^{t170}$) $>$ 0,
 1.0 VECBOS events are predicted
compared to the observation of 6 data events.
The excess with respect to the QCD
 prediction at ln(L$_{rel}^{t170}$) $>$ 0 already observed for
$W+ \geq 3$ jet events
in Figure 10(b), is
therefore made relatively more pronounced by the requirement of a fourth jet,
 showing a
positive correlation between the ln(L$_{rel}^{t170}$) $>$ 0
and four-jet signature.
 The four--jet topology and b-tags are also strongly correlated.
 In three of the 6 four-jet events we find an SVX tag.
The distribution in ln(L$_{rel}^{t170}$) of these 3 events is shown in
 Figure 16(b).  In the absence of top quark, 0.15 background
 tags are predicted for the dominant direct $W +$ jet production,
distributed as
shown in the figure.\\
\indent A similar picture emerges for the SLT tag algorithm, since
 3 SLT tags are identified
in the $W+4$ jet sample.
In conclusion,
6 out of the 8 events at ln(L$_{rel}^{t170}$) $>$ 0 of Fig.  10(b) have
at least one
additional jet, and 5 of them are b--tagged.
This is very unlikely to be due to background, and is more consistent
with $t \bar t$ events. \\
\indent Using the methods described in Ref.~\cite{prd}
we have computed the top quark mass for the subset of
events of the signal and control sample with exclusively
four jets, by requiring $E_{T4}$ $>$ 15 GeV and E$_{T5}$ $<$ 10 GeV.
Four events fulfill this requirement: three belong to the signal sample
 and one to the control sample.
 The three of the signal sample (all at ln(L$_{rel}^{t170}$) $>$ 0) are
 in common with the $W +$ jet event sample of Ref.~\cite{prd})   and
 are among the 7 tagged events
used in~\cite{prd} for the derivation of the top quark mass.\\
\indent The masses
 of these 4 events are
in the range 161 $\pm$ 11 GeV/c$^2$ to 172 $\pm$ 11 GeV/c$^2$, lower on average
but consistent with the result of M$_{top}$ = 174 $\pm$ 10 $^{+13}_{-12}$
reported in Ref.~\cite{prd}. We have compared their mass distribution
 with the distributions expected for $t\bar t$ and direct
$W + 4$ jet events. The expected distributions for top quark and for direct $
W +$ jet
production are appreciably different. Within the very poor statistics,
the distribution of the data events, shown in Figure 17,
favours the $t\bar t$ hypothesis
over QCD.

\begin{figure} [hbt]
\mbox{
\centering
\epsfxsize 14.cm
\epsfysize 14.cm
\epsffile[0 0 567 567]{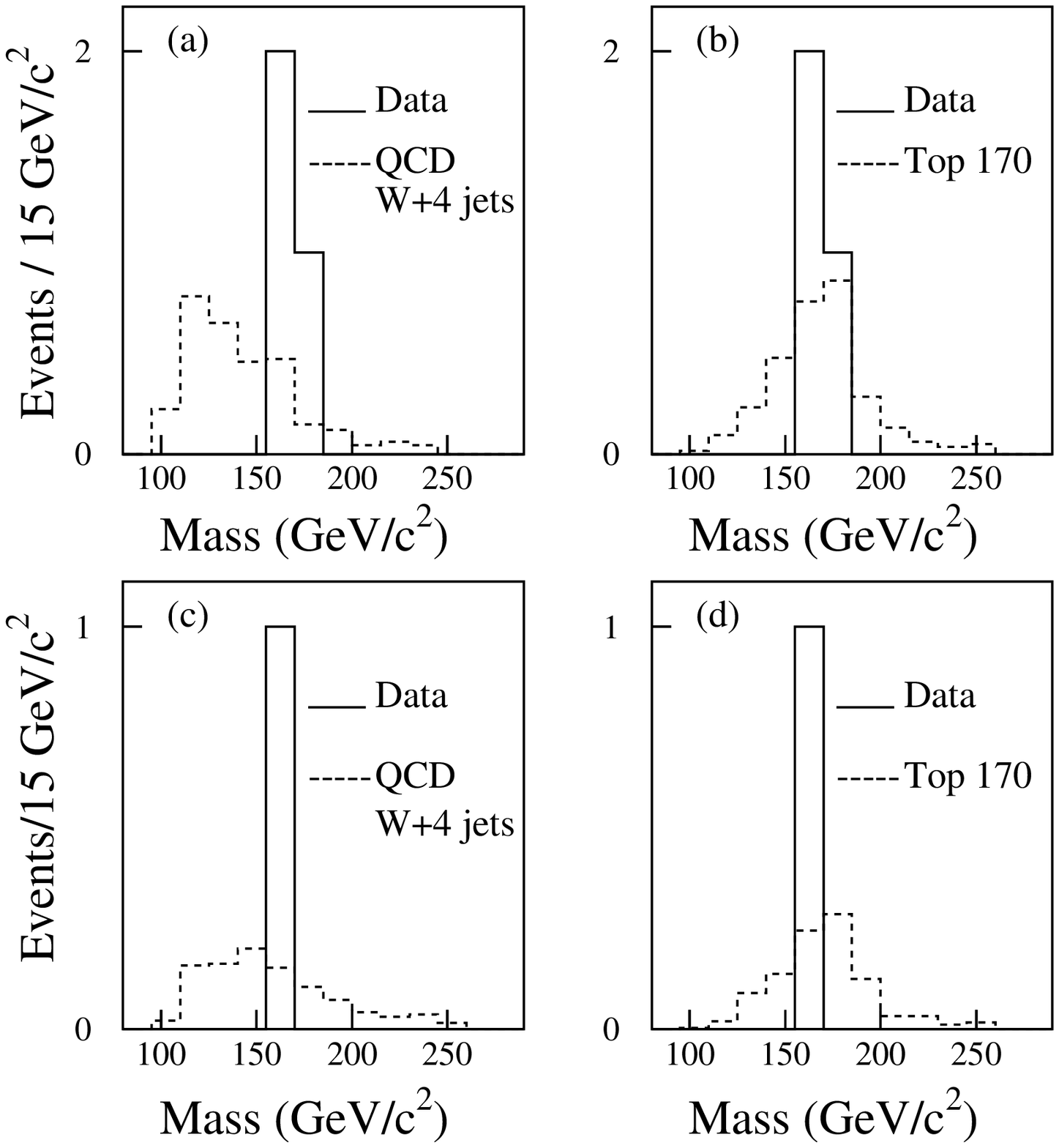}}
\caption{ Distribution of the expected mass for
VECBOS Monte Carlo events, analyzed as they were $t \bar t$ events
(dashed histogram)
(a) for the signal sample and (c) for the control sample cuts.
 Distribution of the preferred mass for $t \bar t$
(M$_{top}$ = 170 GeV/c$^2$) Monte Carlo events, analyzed as $t \bar t$ (b)
for the signal sample and (d) for the control sample cuts. The 4 events
 which allow the mass reconstruction are shown as solid histograms.}
\label{mass}
\end{figure}

\section{Conclusions}

\hspace{0.5cm} The kinematics of a sample of 49
$W+\geq 3$ jet events was compared
with the theoretical expectations for
 direct  $W+$jet production and $t \bar{t}$ quark pair production.
 It is determined whether a given $W+\geq 3$ jet event fits better
the expectations of direct $W +$ jet production as predicted by
the VECBOS QCD Monte Carlo or top quark as predicted
by the HERWIG Monte Carlo.
 The VECBOS predictions for $W +\geq$ 2 central jets  and
 $Z + \geq$ 3 jet production agree well with the observed data.
 A subsample of
$W+\geq 3$ jet events (``control sample'') that should be enriched in
 direct $W$ production events relative to top quark has been defined.
 VECBOS also gives a good description of
the observed jet E$_T$ distributions for this sample.
 A separate subsample (``signal sample'') is defined with the requirement
that the three leading jets be central. It
should be enriched in $t \bar{t}$ events relative to direct $W + $
jet events which form the main background.
 This signal sample contains  14 events.
The jet E$_T$ distributions for these events
are unusually hard and not well described by the expectations from QCD and
other backgrounds.
 By means of a suitable variable, ln(L$_{rel}^{t170}$), events that are
kinematically more top quark--like
can be selected as those events with ln(L$_{rel}^{t170}$) $>$ 0.
We observe 8 such events, while we expect 1.7
from non-top quark processes.
{}From a statistical analysis, which takes into account the
systematic errors, we have derived a probability
of 0.8\% for this excess to be due entirely to
background fluctuations. The analysis was repeated for
a number of different selection cuts defining the signal sample,
and in the worst case a
probability for such a fluctuation as large as 1.9\% was found.
A two
component fit to the data that includes contributions from
a 170 GeV/c$^{2}$ mass top quark and from QCD and other backgrounds
gives a good description of the observed jet E$_T$ distributions,
and yields a \TTbar\ production cross section of
11.6$^{+7.0}_{-5.7}$ $^{+3.2}_{-2.0}$ pb, consistent with the findings
of Ref.~\cite{prd}. A similar two component fit to the background enriched
 control sample yields a cross section which is 1 sigma below this value,
 and statistically consistent with zero.\\
\indent  With a secondary vertex $b$--tag algorithm (SVX) we
find evidence for bottom quark decay in four of
the 14 events in the signal enriched sample.
 If the 14 events contained no contribution from top quark,
 only 0.58 events with such a secondary vertex $b$--tag are expected.
The probability for four events to be tagged  due to a statistical
fluctuation is 0.4\%. Similarly, this same event sample of 14 events
contains 4 soft lepton tags (SLT) with an expected background of 1.2 events.
The probability for four events to be tagged  due to a statistical
fluctuation is about 4\% in this case.

 Additional information on the nature of the
events at ln(L$_{rel}^{t170}$) $>$ 0 was obtained from their large probability
of containing
a fourth jet. In the signal sample, out of a total of 8
events at ln(L$_{rel}^{t170}$) $>$ 0,
 there are 6 four-jet events
and 5 of them are b--tagged.
  Assuming that b--tags are indicative of $t \bar{t}$ pairs, one can argue
that the events at ln(L$_{rel}^{t170}$) $>$ 0 show an increased top quark
 purity when the kinematic cuts are made more stringent (a 4th jet is
required).
We note that
5 out of 6 b-tagged events
of the signal sample listed in Table 5 are in common with the b--tagged
sample of
Ref.~\cite{prd}.
This shows that, although the primary event sample selected in this
analysis overlaps
only in part with the W+jet sample of Ref.~\cite{prd} (25 events in common),
the two analysis strategies have isolated the same physics process.
The evidence for top quark reported in \cite{prd} was derived only on the
basis of the observed excess of di-leptons and b-tags.
 The observation of a top quark--like component in the ln(L$_{rel}^{170}$)
distribution reported here provides additional evidence,
independent of that provided by the counting experiments
reported in Ref.~\cite{prd}, that our data contains
a fraction of events more consistent with the decays of top quarks of mass
around 170 GeV/$c^2$ than with the $W +$ jet background.
\section{Acknowledgments}
We thank the Fermilab staff and the technical staffs of the
participating institutions for their vital contributions.
 We also thank Walter Giele for advice and many helpful suggestions
regarding $W$+jets and the VECBOS Monte Carlo program.
This work is supported by the U.S. Department of
Energy and the National Science Foundation; the Italian Istituto
Nazionale di Fisica Nucleare; the Ministry of Science, Culture,
and Education of Japan; the Natural Sciences and Engineering Council of
Canada; the A. P. Sloan Foundation; and the Alexander von Humboldt--Stiftung.
\clearpage



\begin{thebibliography}{99}
\bibitem{eta}  Pseudorapidity, $\eta$, is defined as
$\eta$=-ln(tan($\theta$/2))
 where $\theta$ is the polar angle in spherical coordinates, measured from the
proton beam axis assuming a z--vertex position of zero.
\end{thebibliography}
\end{document}